\documentclass[graybox]{svmult}
\usepackage{type1cm}        % activate if the above 3 fonts are
\usepackage{makeidx}         % allows index generation
\usepackage{graphicx}        % standard LaTeX graphics tool
\usepackage{multicol}        % used for the two-column index
\usepackage[bottom]{footmisc}% places footnotes at page bottom
\usepackage{rotating}
\usepackage{lscape}
\usepackage{newtxtext}       % 
\usepackage{newtxmath}       % selects Times Roman as basic font
\usepackage{subfiles} % Best loaded last in the preamble
\usepackage{newtxmath}       % selects Times Roman as basic font
\usepackage[numbers]{natbib}% For \citep and related citation commands
\makeindex             % used for the subject index
\usepackage[colorlinks=true,
            citecolor=blue,
            linkcolor=blue,
            urlcolor=blue]{hyperref}

\begin{document}

\title*{Observations of X-ray Quasi-Periodic Eruptions}
\author{T. Wevers, J. Chakraborty$^{\dagger}$, E. Quintin$^{\dagger}$, M. Zaja\v{c}ek, and M. Giustini}
\authorrunning{Wevers, Chakraborty, Quintin, Zaja\v{c}ek and Giustini}
\institute{Thomas Wevers \at Astrophysics \& Space Center, Schmidt Sciences, New York, NY 10011, USA
\and Joheen Chakraborty \at Department of Physics \& Kavli Institute for Astrophysics and Space Research, Massachusetts Institute of Technology, Cambridge, MA 02139, USA 
\and Erwan Quintin \at European Space Agency (ESA), European Space Astronomy Centre (ESAC), Camino Bajo del Castillo s/n, 28692 Villanueva de la Cañada, Madrid, Spain 
\and Michal Zaja\v{c}ek \at Department of Theoretical Physics and Astrophysics, Masaryk University, Kotlářská 2, 611 37 Brno, Czech Republic 
\and Margherita Giustini \at Centro de Astrobiologia (CAB), CSIC-INTA, Camino Bajo del Castillo s/n, Campus ESAC, 28692, Villanueva de la Cañada, Madrid, Spain 
\and $^{\dagger}$ These authors contributed equally.}
\maketitle

\abstract{Quasi-periodic eruptions (QPEs) are a novel class of repeating nuclear transients, discovered exclusively in the X-ray band to date. Since their initial discovery in 2019, the QPE sample has grown to 13 sources, exhibiting large amplitude, quasi-regular eruptive variability patterns that are distinct from previously known modes of massive black hole variability. 
In this chapter, we provide a comprehensive overview of their observational characteristics. We review the X-ray spectral and timing properties of QPE eruptions, their long-term evolution, as well as the underlying quiescent emission, which is well described by thermally dominated, compact accretion disks. We discuss population-level emerging trends and selection biases, and present an updated census of their host galaxy properties. We also highlight the growing body of evidence pointing to strong connections between QPEs and tidal disruption events. Finally, we briefly summarize the key observational constraints on proposed QPE model interpretations, before looking ahead to the observational challenges and opportunities that will shape future progress in understanding this emerging population of nuclear transients.}

% \section{Introduction}
% \subfile{1_introduction.tex}

% \section{Discovery and Historical Development}

\setcounter{section}{0}
\section{Discovery and historical development}
X-ray quasi-periodic eruptions
%\footnote{Historical note: the name ``quasi-periodic eruptions'' was suggested to G. Miniutti by A. Laor while discussing the extraordinary properties of the phenomenon.} (
(QPEs) were serendipitously discovered in the nucleus of the galaxy GSN 069 during a 12-hour XMM-Newton observation performed at the end of 2018. This observation revealed two sharp flares in the X-ray light curve of the source. These flares were subsequently confirmed not to be random events in a dedicated, longer (about 36 hours) XMM-Newton observation in January 2019, which detected five luminous flares lasting about one hour and recurring quasi-periodically roughly every nine hours \cite{2019Natur.573..381M}. 

The phenomenon is unlike any known supermassive black hole (SMBH) variability. The galaxy GSN 069 had been monitored since 2010, when it was observed by the XMM-Newton Slew survey \cite{2008A&A...480..611S} to display an X-ray photon flux several orders of magnitude higher than the upper limits provided by the ROentgen SATellite (ROSAT, \cite{Aschenbach1981}) in the 1990s. Many XMM-Newton \cite{2001A&A...365L...1J} and Swift \cite{2004ApJ...611.1005G} observations followed, showing a slow decay in the X-ray flux over more than a decade, which may suggest a TDE of an evolved star (e.g., a red giant; \cite{2021ApJ...920L..25S, 2023A&A...670A..93M}). 
This evolution was intriguing because of the soft X-ray spectrum lacking the power-law emission characteristic of active galactic nuclei (AGNs), with negligible amounts of (neutral) absorption along the line of sight, while only narrow emission lines were present in an optical spectrum of the host galaxy \cite{2013MNRAS.433.1764M}. As such, it caught the attention of astronomers trying to understand the structure of accreting SMBHs.
A more exhaustive study showed that XMM-Newton performed a 24 hour pointed observation of GSN 069 in 2014 (i.e. prior to the QPE discovery). This observation was long enough for QPEs to be detected, if they were present; their absence may have been intrinsic or due to observational biases (as will be discussed later). 

After the discovery of QPEs in GSN 069, an archival search rapidly identified a second QPE-hosting galaxy, RX J1301.9+2747 \cite{2020A&A...636L...2G}, a post-starburst galaxy in the outskirts of the Coma Cluster. RX J1301.9+2747 has displayed peculiar X-ray variability since the 1990s, with a sharp decline in X-ray emission detected by ROSAT \cite{2000MNRAS.318..309D}. An XMM-Newton observation in 2000 revealed one-and-a-half sharp X-ray flares \cite{2013ApJ...768..167S, 2017ApJ...837....3S}, and an XMM-Newton DDT ToO \footnote{Director's Discretionary Time Target of Opportunity, granted by the XMM-Newton Project Scientist at the time, N. Schartel.} observation performed in 2019 confirmed the QPE-like nature of RX J1301.9+2747 with three sharp flares lasting about half an hour and separated by about five and three hours \cite{2020A&A...636L...2G}, and spectral properties very similar to those of the QPEs seen in GSN 069. 
RX J1301.9+274 is the source with the longest observed QPE lifetime to date, likely lasting for at least 22 years \cite{2024A&A...692A..15G}. 

In 2019, the Russian-German Spectrum-Roentgen-Gamma mission with the eROSITA instrument on board was launched and started a sensitive, multi-epoch survey of the X-ray sky \cite{2021A&A...647A...1P}.
The discovery of QPEs prompted the design of algorithms to discover more sources with similar characteristics. This effort was very successful: using the $\sim$2.5 years of data of the eROSITA all sky survey, five QPE sources were discovered (named eRO-QPE1, eRO-QPE2, ..., eRO-QPE5; see \cite{2021Natur.592..704A, 2024A&A...684A..64A, 2025ApJ...989...13A}). This is a substantial fraction of the known QPE sample to date, and the increased sample size has helped to enable the first statistical studies of QPE sources as a population.

The first two QPE sources found by eROSITA were quite dissimilar in their properties: while eRO-QPE2 was akin to the already discovered GSN 069 and RX J1301.9+2747 in terms of QPE duration and time separation (or recurrence time: the time between two consecutive QPEs), eRO-QPE1 showed much longer duration (about eight hours) and time separation between flares (about eighteen hours). Indeed, even the longest uninterrupted XMM-Newton observation is unable to clarify the QPE nature of the source eRO-QPE1, and only continuous high-cadence monitoring with NICER reveals the quasi-periodic nature of its flares.

As more sources were discovered, observational evidence for a connection between QPEs and TDEs started emerging, both from their host galaxies \cite{2022A&A...659L...2W} and their long-term evolution \cite{2023A&A...670A..93M, 2023ApJ...957...34L, 2024arXiv241104592S, 2024ApJ...970L..23W, 2026A&A...709A.147M}. This prompted astronomers to start targeted X-ray monitoring observations of optical TDEs months-years after peak. The first spectacular confirmation of the QPE-TDE physical connection came with AT2019qiz \cite{2024Natur.634..804N}, where quasi-periodic eruptions were discovered serendipitously in late-time Chandra follow-up observations $\sim$4 years after the optical outburst. This method has proven to be very efficient and has since been expanded to target TDEs, TDE candidates and large amplitude nuclear flares at late times, leading to the discovery of three new QPE sources to date: AT2022upj \cite{2025ApJ...983L..39C}, ZTF19acnskyy \cite{2025NatAs...9..895H}, and the eROSITA-discovered TDE candidate eRASStJ2344 \cite{2026arXiv260203932B}. Early statistical surveys estimate that around $9^{+9}_{-5}$\% of optically-selected TDEs show QPEs within five years post-peak, providing a rough estimate for the efficacy of TDE monitoring campaigns to discover further QPEs.

In parallel, more sophisticated and thorough searches of the X-ray mission archives have yielded two `QPE candidates': XMMSL1 J024916.6-041244 (J0249, \cite{2021ApJ...921L..40C}) and AT2019vcb (Tormund, \cite{2023A&A...675A.152Q}). 
In the former source, one and a half flares were observed in an archival (2006) XMM-Newton observation of the TDE candidate J0249, which showed strong spectral similarities to other QPEs as well as a long-term X-ray evolution similar to GSN 069. Repeating flares were not, however, detected in more recent XMM-Newton DDT observations in 2021. Nevertheless, the spectral characteristics of the 1.5 flares detected in 2006 make J0249 generally accepted as a QPE source. 

\begin{landscape}
\begin{figure*}
    \centering
    \includegraphics[width=\linewidth]{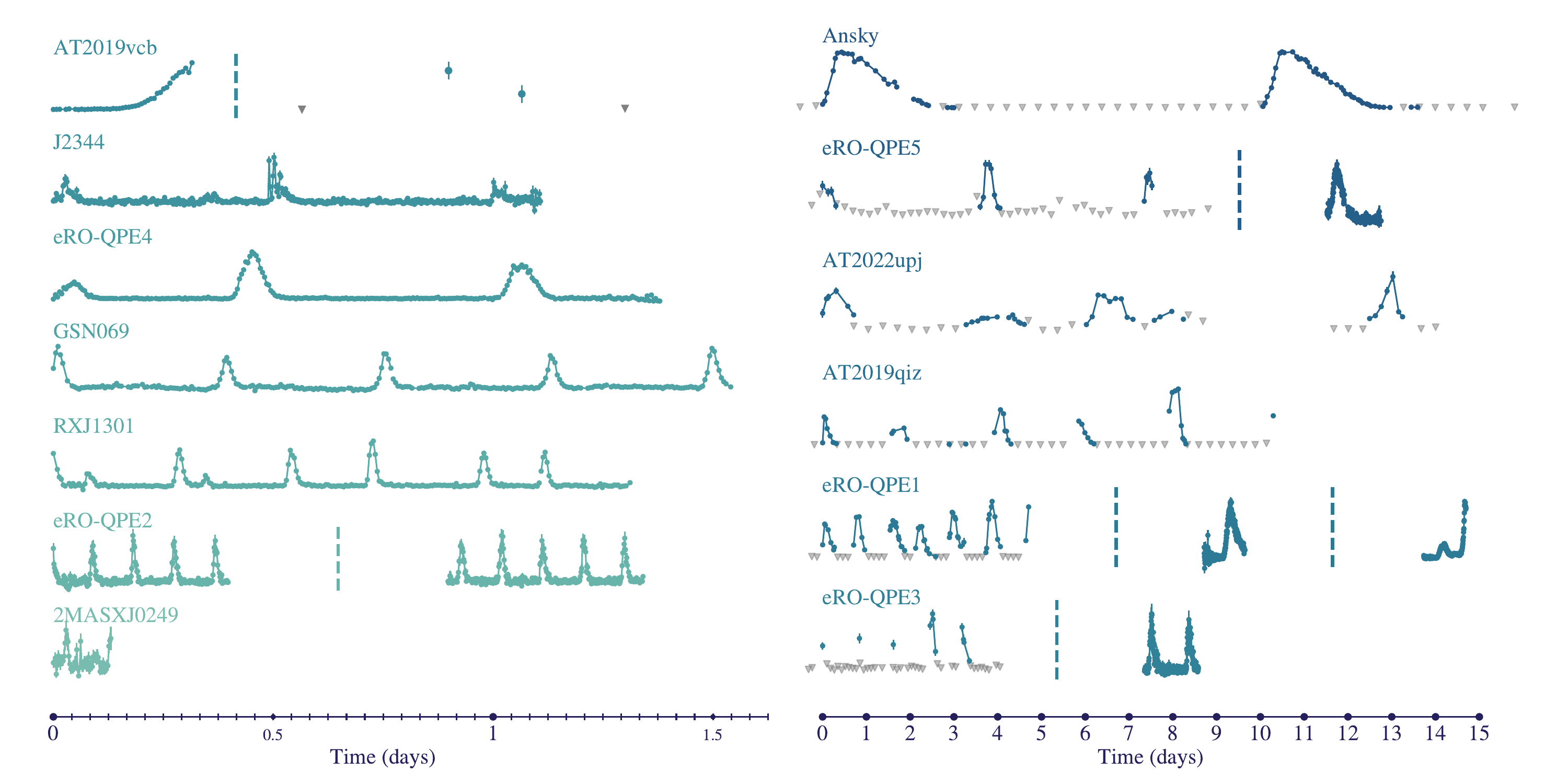}
    \caption{X-ray lightcurves of the 13 QPE sources known at the time of writing, showing their general similarities but also the heterogeneity in recurrence time, stability, and eruption profiles. Dots represent detections, while gray triangles represent upper limits. The data corresponds to publicly available XMM-Newton observations for the shorter QPEs (left column) and NICER data for the longer QPEs (right column). Different observations of a given object are separated by a dashed line to indicate time gaps.}
    \label{fig:all_lightcurves}
\end{figure*}
\end{landscape}

In AT2019vcb, the rise of a flare very similar to the ones observed in eRO-QPE1 in terms of intensity, duration, and spectral properties was captured in an XMM-Newton observation. This flare was observed six months after the optical peak of a TDE \cite{2023A&A...675A.152Q}; however, the observation ended mid-flare. Several months later, eROSITA observations confirmed the QPE nature of this source, dubbed Tormund/AT2019vcb \cite{2025MNRAS.540...30B}.
In Figure \ref{fig:all_lightcurves} we show a compilation of the lightcurves of all known QPE sources to date, highlighting the wide range of recurrence times, flare profiles, and amplitudes.

% \bibliographystyle{style/spphys.bst}
% \bibliography{references.bib}

% \section{Observational Properties}

\setcounter{section}{2}
\section{Observational properties}
\label{sec:properties}
% \textbf{Some intro written by MZ, feel free to add, modify}\\
% \subsection{Working definition}
At the time of writing, there is not yet an exact definition of what constitutes a  QPE source. We therefore start with a working definition of the observational properties that are broadly used by the community to classify a source as a QPE: \\

{\bf Definition}\\
Quasi-periodic eruptions are recurring bursts of thermal X-ray radiation from the nuclei of galaxies that occur at almost regular intervals, with each eruption shorter in duration than the time between successive events. Their soft X-ray spectral and timing evolution traces a counterclockwise loop in the temperature–luminosity (or hardness ratio–count rate) plane: the eruption temperature rises rapidly before reaching a peak in both luminosity and temperature, followed by a decline in brightness at a cooler temperature than during the rise (a behavior known as temperature hysteresis).\\

%One of the defining observational properties of QPEs, in comparison with other active and quiescent galactic nuclei, are recurrent, quasi-periodic bursts or eruptions 
To date, such regular recurring flares on hours-to-days timescales around SMBHs have only been observed in the soft X-ray band ($ 0.2 < E <2$ keV), with durations ranging from a few tens of minutes to a few days, and recurrence times from $\sim 2.5$ hours to 12 days (see e.g. \cite{2019Natur.573..381M,2021Natur.592..704A,2024A&A...684A..64A}, and Figure \ref{fig:all_lightcurves}). 
With the exception of a delayed and temporally smeared UV brightening correlated with the eruptions has been reported in one source (ZTF19acnskyy/Ansky; \cite{2026ApJ..1000L..57G}), no prompt eruption counterpart at other wavelengths has yet been established.

The eruptions have amplitudes of one to two orders of magnitude (in the observed band) compared to the out-of-eruption emission, and a duty cycle, defined as the ratio of the eruption duration to the recurrence timescale $D=\tau_{\rm erupt}/\tau_{\rm rec}$, of $\sim$0.1--0.3. 
These properties are inconsistent with the stochastic red-noise variability characteristic of most AGN, and hence represent a new phenomenological class of SMBH variability. 

In this Section we describe the properties of the eruptions for which the systems are known as well as the observed quiescent emission. These two components provide complementary insights into the nature of the phenomenon.

% In the X-ray band, the emission is soft and is consistent with the thermal emission both during the quiescence as well as during eruptions. The X-ray spectra can be characterized by a (blackbody) temperature of $kT_{\rm QPE}\sim 50\,{\rm eV}$ ($T\sim 6 \times 10^5\,{\rm K}$) at the beginning and the end of the eruption. Close to the eruption peak, there is a clear trend of spectral hardening, with typical temperature of order $kT_{\rm QPE}\sim 100\,{\rm eV}$ ($T\sim 1.2\times 10^6\,{\rm K}$) close to the eruption peak. The flare emission generally lacks the hard X-ray power-law tail beyond 1-2 keV. The peak X-ray luminosities are in the range $\sim 10^{42}-10^{44}\,{\rm erg\,s^{-1}}$, which is 1--2 orders of magnitude larger than the quiescent luminosity between the eruptions.

\begin{table*}[h]
\centering
\resizebox{\columnwidth}{!}{
\begin{tabular}{cccccccccccccc}
    \hline\hline \\[-0.1cm]
QPE & RA & Dec & Eruption duration & Recurrence time & Peak X-ray Luminosity & Quiescent X-ray Luminosity & Peak Temp. & Quiescent Temp.& z \\[0.1cm]
                  && & (ks)            & (ks)              &     ($\times10^{42}$~erg~s$^{-1}$)               &($\times10^{42}$~erg~s$^{-1}$)  &   eV & eV   \\[0.2cm] \hline \\
GSN 069\textsuperscript{1} & 01 19 08.66 & --34 11 30.54  & $\sim3.5$  & $\sim [20-33]$ & $\sim [1-5]$ & $\sim [2-8]$ &$\sim[75-105]$& $\sim[45-60]$& 0.018 \\
RXJ1301\textsuperscript{2} & 13 02 00.14 & 27 46 57.83  & $\sim[1.5-3]$  & $\sim [5-20]$  & $\sim [1.0-1.6]$ & $\sim [0.1-0.2]$ &$\sim[100-200]$&$\sim[50-60]$& 0.024 \\
2MASXJ0249\textsuperscript{3} & 02 49 17.32 & --04 12 52.13 & $\sim1$ & $\sim9$ & $\sim0.3$ & $\sim0.16$ &$165\pm20$ & $75\pm7$ & 0.019\\
eRO-QPE1\textsuperscript{4} & 02 31 47.16 & --10 20 10.97  & $\sim[20-30]$  & $\sim[18-80]$ & $\sim[3-20]$ & $0.2\pm0.06$ &$\sim [150-260]$&$130\pm30$& 0.050 \\
eRO-QPE2\textsuperscript{4} & 02 34 48.70 & --44 19 32.78 & $\sim1.7$  & $\sim[8-9]$  & $1.2\pm0.2$ & $0.1\pm0.01$ &$220\pm20$&$75\pm5$& 0.018 \\
eRO-QPE3\textsuperscript{5} & 14 00 53.33 & --28 46 01.21 & $\sim12$ & $\sim[60-70]$  & $\sim[0.3-3]$ & $\sim[3\times10^{-2}-3]$ &$110\pm20$&$46\pm20$& 0.024 \\
eRO-QPE4\textsuperscript{5} & 04 45 33.81 & --10 12 04.76 & $\sim9$  & $\sim[30-50]$  & $\sim[7-12]$& $1.7\pm0.3$ &$120\pm2$&$43\pm2$& 0.044 \\
eRO-QPE5\textsuperscript{6} & 03 25 43.21 & --45 12 45.12  &  $\sim70$ & $\sim320$ & $30.0\pm0.3$ & $1.8\pm0.7$ &$110\pm2$&$37\pm10$& 0.116 \\
AT2019vcb ("Tormund")\textsuperscript{7} & 12 38 56.38 & 33 09 57.28 & $>$15  & $<1000$ & $119\pm5$ & $3.2\pm1.0$ &$114\pm3$&$53\pm8$& 0.088 \\
AT2019qiz\textsuperscript{8} & 04 46 37.880 & --10 13 34.92  & $\sim$35  & $\sim[140-200]$  & $\sim[5-10]$ & $\sim0.1$ & $109\pm1$&$67\pm10$&0.015\\
AT2022upj\textsuperscript{9} & 00 23 56.85 & --14 25 23.22 & $\sim[25-75]$ & $\sim[40-300]$  & $\sim[3-10]$ & $2.6\pm0.3$ &$210\pm20$&$211\pm30$& 0.054\\
ZTF19acnskyy (``Ansky'')\textsuperscript{10} & 13 35 19.92 & 07 28 07.54 & $\sim [130-260]$ & $\sim [380-1000]$ & $\sim[10-40]$ & $\sim0.02$ &$\sim[90-110]$&$50\pm10$& 0.024 \\
eRASSt J2344\textsuperscript{11} & 23 44 02.95 & --35 26 41.67  & $\sim4$  & $\sim[40-45]$  & $\sim10$ & $5.4\pm0.5$ &$\sim[80-100]$&$56\pm3$& 0.10\\[0.2cm]
 \hline\hline \\[-0.1cm]
\end{tabular}
}
\caption{Summary of the observational properties of QPEs. Luminosities are given in the 0.2-2 keV band. We emphasize that the properties of QPEs generally evolve over time, and as such the values in this table are meant to be indicative of the range of possible values. In particular, for individual sources ranges of values taken across epochs are indicated within brackets; for the sake of readability, we do not indicate individual errors on these measurements, as they are smaller than the observed range of values. References: \textsuperscript{1} \cite{2019Natur.573..381M}, \textsuperscript{2}\cite{2020A&A...636L...2G},\textsuperscript{3}\cite{2021ApJ...921L..40C}, \textsuperscript{4}\cite{2021Natur.592..704A,2022A&A...662A..49A},\textsuperscript{5}\cite{2024A&A...684A..64A}, \textsuperscript{6}\cite{2025ApJ...989...13A},\textsuperscript{7}\cite{2023A&A...675A.152Q}, \textsuperscript{8}\cite{2024Natur.634..804N},\textsuperscript{9}\cite{2025ApJ...983L..39C},\textsuperscript{10}\cite{2025NatAs...9..895H,2025A&A...703A.263H}, \textsuperscript{11}\cite{2026arXiv260203932B}}
\label{tab:Population}
\end{table*}

\subsection{Eruptions}
\label{sec:eruptions}

\subsubsection{Spectral evolution}
The spectra of the eruptions are thermal, with the X-ray continuum well described by the Wien tail of a $\sim 60-250$ eV blackbody \cite{2019Natur.573..381M,2020A&A...636L...2G,2021Natur.592..704A,2021ApJ...921L..40C,2023A&A...675A.152Q,2024A&A...684A..64A,2024Natur.634..804N,2025ApJ...983L..39C,2025NatAs...9..895H}. The blackbody temperature changes significantly during the eruptions, with initial values in the range $80-150$ eV, peak temperatures in the range $100-250$ eV, and decline temperatures in the range $60-120$ eV. As a consequence of their spectral properties, most of the photons are emitted in the far UV and soft X-ray bands, typically below $\sim$ 1.5 keV.

\begin{figure*}
    \centering
    \includegraphics[width=\textwidth]{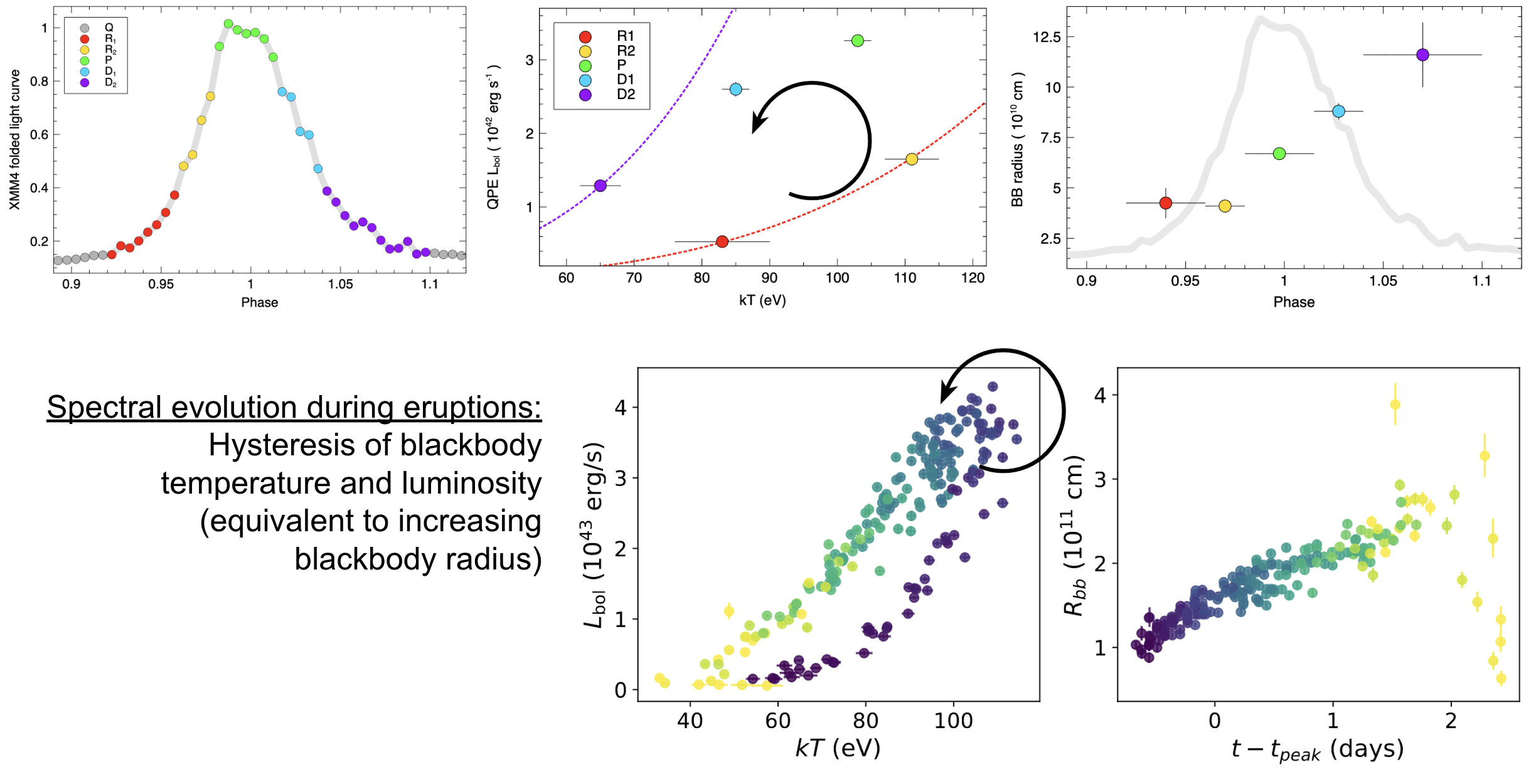}
    \caption{Spectral evolution during the eruptions. \textbf{Top:} flare profile, temperature hysteresis, and inferred blackbody radius in GSN 069. \textbf{Bottom:} same for ZTF19acnskyy. Figures adapted from \cite{2023A&A...670A..93M} and \cite{2025A&A...703A.263H}.}
    \label{fig:spec_evol}
\end{figure*}

A key characteristic of the spectral behavior is a hysteresis loop whereby the evolution of the temperature leads that of the luminosity (Figure~\ref{fig:spec_evol}); in other words, the measured blackbody temperature (or equivalently, the hardness ratio) peaks earlier than the continuum flux, and the rise is hotter/harder than the decline. 
%The coupled temperature and luminosity information can be used to crudely infer a characteristic blackbody emission radius assuming $L\propto R_{\rm BB}^2T^4$. In the case of QPEs, $R_{\rm BB}$ at the start of an eruption is of the order $10^{10}-10^{11}\mathrm{cm}\;(\sim \rm 1\ solar\ radius,\ R_\odot)$, and then increases by a factor of a few over the course of the evolution of the eruption.

The cosmological volume for detecting QPEs with peak luminosities up to 10$^{44}$ erg s$^{-1}$ in principle extends to $z\sim$1 (assuming a 1 ks XMM-Newton limiting depth of 2$\times$10$^{-14}$ erg cm$^{-2}$ s$^{-1}$ or a 3 ks Swift/XRT limiting depth of $\sim$10$^{-13}$ erg cm$^{-2}$ s$^{-1}$). However, an increasing fraction of the photons is shifted out of the observable X-ray band as the source redshift increases.

The handful of sources with long-term monitoring has shown a wide variety of  behaviors. In the cases of GSN 069 and eRO-QPE1, the eruptions show dramatic variability in their overall strength, decreasing over the course of a few years by factors of $4\times$/$90\times$ in luminosity and $1.4\times$/$3\times$ in temperature, respectively \cite{2023A&A...674L...1M,2024ApJ...965...12C};the QPEs have decreased in amplitude by a factor of $10\times$ in two years for eRO-QPE3 \cite{2024A&A...684A..64A}. Interestingly, $k_BT_{\rm peak}$ and $L_{\rm peak}$ in both sources changed together while roughly following $L_{\rm peak}\propto T_{\rm peak}^4$, suggesting an underlying process with a characteristic emission radius being preserved. 

In the cases of RX J1301 and eRO-QPE2, no such dramatic changes in temperature \textit{or} luminosity were seen over comparable or longer timescales, with the burst spectra appearing remarkably stable over a three year monitoring campaign (eRO-QPE2; \cite{2024A&A...690A..80A}) and, in the case of RX J1301, over the $\sim$two decades for which archival data is available \cite{2024A&A...692A..15G}.
% \newline

\subsubsection{Temporal Characteristics}
The observed recurrence times of QPEs span about two orders of magnitude across the known sample (see Fig. \ref{fig:all_lightcurves}), ranging from 2.5 hours up to $\gtrsim$12 days. The eruptions tend to recur at approximately consistent intervals, maintaining a characteristic quasi-period to within a few tens of percent over dozens of cycles. However, strictly constant spacing is not maintained indefinitely: small cycle-to-cycle variations are present, and some sources show considerable variability in their recurrence patterns. 

For example, RXJ1301 has an average recurrence time of 4 hours but with spacings as short as 1 hour and as long as 6 hours \cite{2024A&A...692A..15G}, while in eRO-QPE1 there have been phases where subsequent peaks overlap each other temporally, and display variations of a factor of 2--5 in peak luminosity \cite{2022A&A...662A..49A}. In addition, a few sources show an alternating long-short pattern in their flares (e.g. \cite{2019Natur.573..381M,2021Natur.592..704A}). This pattern can be accompanied by a corresponding variation in the eruption peak luminosity: for example, a longer wait time is followed by a stronger eruption for RXJ1301 \cite{2024A&A...692A..15G}. Such a trend is not usually seen in sources without long-short alternation (e.g. \cite{2024ApJ...965...12C}).

The eruptions are relatively short compared to the recurrence times; the shortest outbursts last for $\sim$20 minutes, while the longest known flares have a duration of 3 days. 
%The typical active duty cycle of QPEs is 10-30 per cent, i.e. these sources spend most of their time in quiescence. 
The burst profiles tend to be asymmetric: most QPEs show a rapid rise to peak, followed by a more gradual decay back to quiescence. Some eruptions are roughly symmetric or display sub-structure (e.g. a plateau or multiple peaks \cite{2022A&A...662A..49A}). A commonly used parametric model which can successfully describe the eruption profiles of several QPEs was introduced in \cite{2022A&A...662A..49A}:
\begin{equation*}
    L(t) = \begin{cases}
        L_{\rm peak}\lambda e^{\tau_1/(t_{\rm peak}-t_{\rm as}-t)} & \mathrm{if\;}t\leq t_{\rm peak}\\
        L_{\rm peak}e^{-(t-t_{\rm peak})/\tau_2} & \mathrm{if\;}t>t_{\rm peak}
    \end{cases}
\end{equation*}
where $L_{\rm peak}$ is the flare amplitude, $t_{\rm peak}$ is the peak timing, and $\tau_1$/$\tau_2$ are the rise/decay $e$-folding times. $\lambda\equiv \exp(\sqrt{\tau_1\tau_2})$ and $t_{\rm as}\equiv\sqrt{\tau_1\tau_2}$ are derived parameters setting the normalization and asymptote time, and are not free parameters. Despite the wide range in recurrence times, patterns and eruption profiles, QPEs have very similar peak luminosities in the 0.3--2 keV band, L$_X$ $\approx 10^{41-43}$ erg s$^{-1}$.  Typical amplitudes compared to the quiescent emission range from 10--100.

Over longer timescales (months to years), QPE timing properties can evolve significantly, indicating that these eruptions are a transient phase rather than a permanent quasi-periodic signal.
Some sources have maintained very stable recurrence times and burst profiles over multiple years of monitoring (e.g. eRO-QPE2 \cite{2024A&A...690A..80A, 2024arXiv241100289P}, RX J1301 \cite{2024A&A...692A..15G}). 
Others have shown dramatic changes in their properties; 
%Several QPEs have shown rapid decreases in their flare luminosities, e.g. by a factor of$10\times$ in two years for eRO-QPE3 \cite{2024A&A...684A..64A}. 
in eRO-QPE1 and 2MASXJ0249, the QPEs have altogether shut off after their initial detection. 
QPEs seemingly ceased in GSN 069 for a period of two years, only to return with shorter recurrence times as well as significantly changed intensities and peak temperatures \cite{2023A&A...674L...1M}. This behavior appears to be coupled to changes in the quiescence accretion luminosity, whereby higher accretion rates are associated with lower QPE luminosities. The threshold for the appearance of QPEs was found to be $(0.4\pm 0.2)L_{\rm Edd}$, suggesting that that QPEs may be an accretion rate dependent phenomenon \cite{2023A&A...674L...1M}. Yet another surprising change was observed in ZTF19acnskyy \cite{2025A&A...703A.263H}, where the recurrence time doubled and the eruption luminosity increased by a factor of 4, in addition to the eruption profiles becoming more asymmetric.

The fact that observations over longer baselines continue to reveal new and unexpected behavior motivates dedicated monitoring campaigns of known sources, and highlights the need to increase the sample size in order to fully map out observational behavior. 

%\subsubsection{Constraints on long-term evolution}

%The known QPEs have shown a wide range of behaviors in their long-term evolutions. Some sources have shown extremely stable QPEs, with no evolution in quiescence and QPE peak luminosities, e.g. over a span of five years in eRO-QPE2 \cite{Arcodia2026} and two decades in RX J1301 \cite{2024A&A...692A..15G}. Other sources have shown rapid decreases in their flare luminosities, e.g. by a factor of $>90\times$ in three years for eRO-QPE1 \cite{2024ApJ...965...12C} and $10\times$ in two years for eRO-QPE3 \cite{2024A&A...684A..64A}. This may naturally extend to the QPE candidates 2MASXJ0249 and Tormund, in which follow-up observations did not detect QPEs, possibly because they have faded below quiescence as one would extrapolate from the former emission level. 

%The most complex evolution is seen in GSN 069, which has shown the dimming, disappearance, and subsequent reappearance of QPEs over $\sim$few years. This behaviour appears to be coupled to changes in the quiescence accretion luminosity, whereby higher accretion rates are associated with lower QPE luminosities. The threshold for the appearance of QPEs was found to be $(0.4\pm 0.2)L_{\rm Edd}$ \cite{2023A&A...674L...1M}. The population of known QPEs is thus quite heterogeneous in their evolutionary properties, suggesting variety in the physical mechanisms responsible for the long-term luminosity behavior of QPEs.

\subsubsection{QPE timing variations}

It is noteworthy that both regular and irregular QPEs exhibit super-periodic, quasi-sinusoidal modulation of the eruption times, measured via $O-C$ (observed minus calculated) diagrams with the QPE arrival times. The $O-C$ method is a model-agnostic method to detect super-periodic variations on timescales $\gg t_{\rm rec}$, as would be expected from (for example) precessional motion of the accretion flow. It provides a key independent check of the otherwise entirely model-dependent tests obtained from fitting QPE timings with EMRI trajectory modeling codes \cite{2021ApJ...921L..32X,2024PhRvD.110h3019Z,2024ApJ...965...12C,2025ApJ...992..120C}. 

\begin{figure}
    \centering
    \includegraphics[width=\linewidth]{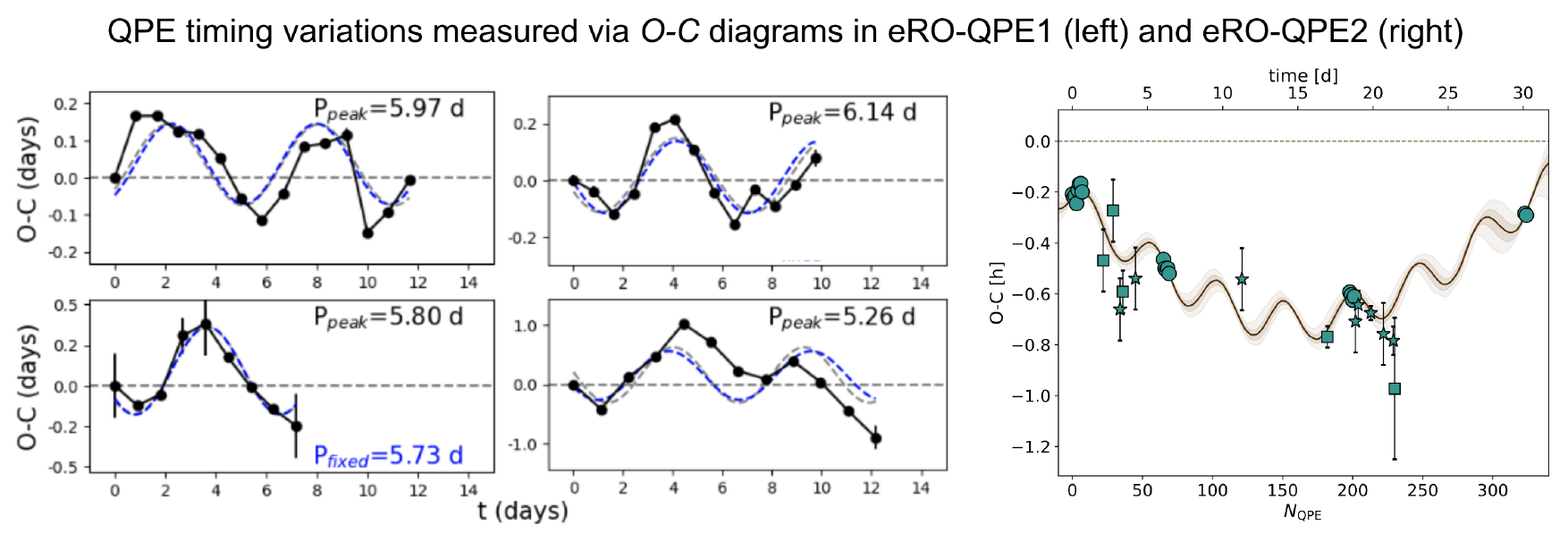}
    \caption{$O-C$ diagrams for eRO-QPE1 (\textbf{left/middle)} and eRO-QPE2 \textbf{(right)} showing evidence for super-periodic oscillations in the QPE times of arrival. Similar evidence has been seen in GSN 069 \cite{2025A&A...693A.179M} and ZTF19acnskyy \cite{Chakraborty2026}. Figures taken from \cite{2024ApJ...965...12C} and \cite{Arcodia2026}.}
    \label{fig:oc}
\end{figure}

$O-C$ diagrams have been constructed for four QPEs at the time of writing, two of which are shown in Fig.~\ref{fig:oc}. In eRO-QPE1, \cite{2024ApJ...965...12C} found that the burst arrival times oscillate approximately sinusoidally in most epochs, with a typical period of $\sim 6t_{\rm rec}$. In eRO-QPE2, \cite{Arcodia2026} found two hierarchically superimposed oscillations of $46t_{\rm rec}$ and $\gtrsim 1000t_{\rm rec}$, and they similarly do not exhibit the predicted anti-correlation. Prior work on eRO-QPE2 also found that it exhibited discrete declines in its recurrence times between 2020-2022 \cite{2024A&A...690A..80A}. Notably, they also found that the super-periodic variations of even and odd bursts do not exhibit the strict anti-correlation predicted by orbiter-disk collision models with two observable QPEs per orbit. 

In GSN 069, \cite{2025A&A...693A.179M} found a similar super-periodic oscillation, with an ambiguous period of either $19$ days or $43-44$ days due to the limited burst sampling, finding the same disagreement with the orbiter-disk collision model. Finally, in ZTF19acnskyy, \cite{Chakraborty2026} found a smooth increase in the burst recurrence times, consistent with a stable period derivative of $\dot{P}\sim 10^{-2}$ over $\gtrsim1$ year of monitoring. The increase can be interpreted as a long-term oscillation of period $11-27$ years, and is also accompanied by a lower-amplitude, shorter-period oscillation with a period of $154-155$ days. The presence of these super-periods, and their hierarchical superposition in eRO-QPE2 and ZTF19acnskyy, is suggestive of multiple physical processes driving the QPE timings, and is a promising avenues for further investigation. 

However, a clear physical interpretation is still lacking, due to the inability of EMRI models to explain the unexpected correlation between even/odd burst timings and the lack of clear predictions from disk instability models. Moreover, it may be unusual in the EMRI scenario that no QPEs have yet exhibited a smooth negative period derivative, and it is very surprising that ZTF19acnskyy has shown a positive $\dot{P}$. These results should be interpreted with some caution, given that the sparse nature and limited time span of the available data can lead to some ambiguity in the $O-C$ results; these preliminary results motivate higher-cadence and longer-baseline monitoring for future QPE timing experiments.

\subsubsection{Absorption lines and outflows}
Evidence for discrete spectral features indicating possible broadened and/or blueshifted absorption lines (usually interpreted as a signature of an outflow) has been reported for several sources. In GSN 069, a broad absorption-like dip near $\sim 0.7$ keV was first noted by \cite{2013MNRAS.433.1764M}, where the authors interpreted the feature as a warm absorber with a column density $N_H\sim10^{22}$ cm$^{-2}$ and an ionization parameter $\log\xi\sim0.4$. At that time, the data quality could only support moderate-resolution spectroscopy with the EPIC CCD detectors of XMM-Newton. Detailed follow-up observations of the source found further evidence of absorption features in XMM-Newton's higher resolution Reflection Grating Spectrometer instrument. These features are best described by a photoionized plasma with $N_H\sim 10^{22}$ cm$^{-2}$, $\log\xi\sim3.9-4.6$, and a blueshifted velocity of $1700-2900$ km s$^{-1}$ \cite{2025ApJ...978...10K}. 

The bulk velocity and column density of the outflow do not change between eruptions and quiescence, but the outflowing material is more highly ionized during the eruptions, as expected if the bursts directly contribute to the plasma ionization state. Given the steady-state nature of the outflow, it is unlikely to be related to the QPE-generating process itself; instead, it is most probably linked to the long-term TDE-like variability observed in GSN 069. Long-lived outflow signatures have also been observed in a number of X-ray bright TDEs \cite{2018MNRAS.474.3593K,2024ApJ...963...75W,2025ApJ...981L..14A}, which are likely related to radiation-pressure driven accretion disk winds. It is plausible that this is also the case in GSN 069.

\begin{figure*}
    \centering
    \includegraphics[width=\textwidth]{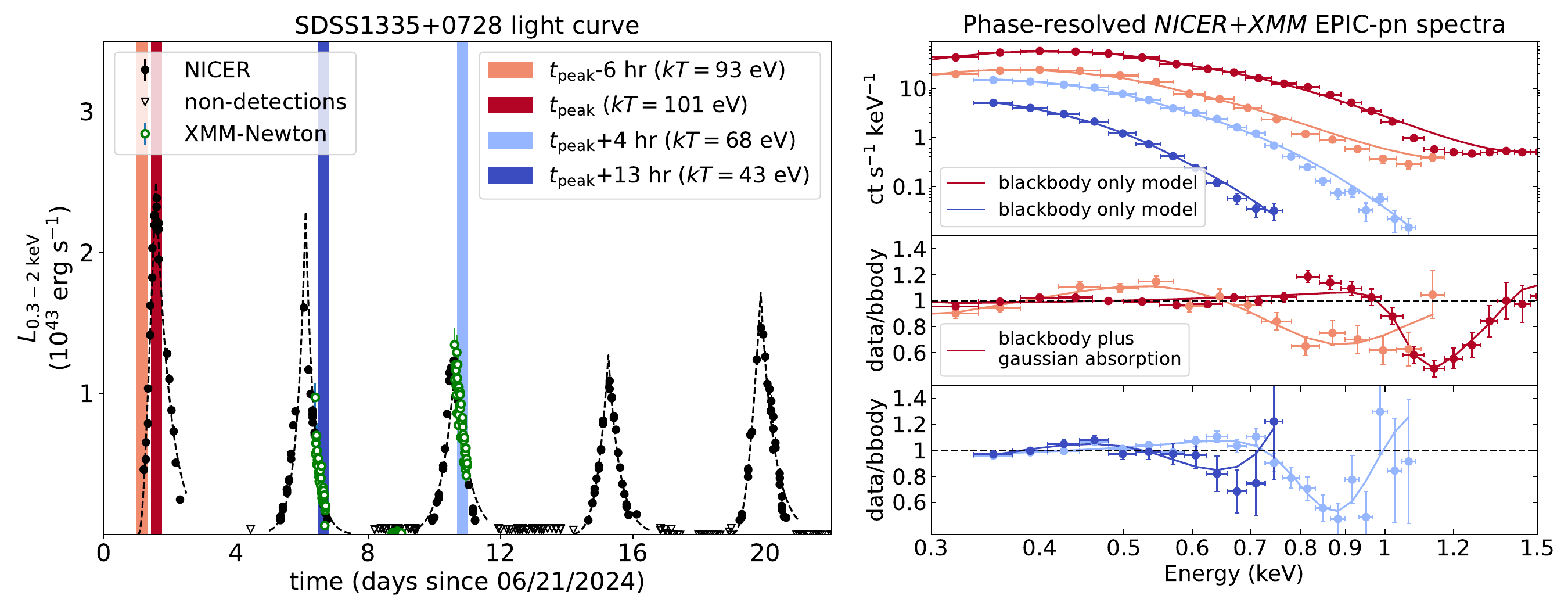}
    \caption{\textbf{Left:} Five consecutive QPEs from ZTF19acnskyy as observed with \textit{NICER} and \textit{XMM-Newton}. Shaded vertical bands represent phases at which spectra were extracted, covering the eruption rise, peak and decline. \textbf{Right:} Phase-resolved spectra fit with a blackbody model, showing time-evolving Gaussian absorption residuals between $\sim 0.6-1.2$ keV. Figure from \cite{2025ApJ...984..124C}.}
    \label{fig:outflow_spectra}
\end{figure*}

A more perplexing case where absorption features have been detected is ZTF19acnskyy \cite{2025NatAs...9..895H}, shown in Figure \ref{fig:outflow_spectra}. Here, the spectral lines show remarkably different characteristics and evolution: the absorption line strength, location, and width are all observed to scale with luminosity and temperature \cite{2025ApJ...984..124C}. Over the course of a single burst, the absorption lines first appear near 0.6 keV during the rise, gradually shifting up toward 1.3 keV, then back down to 0.8 keV, all while maintaining a line width $\sigma\sim0.05-0.15$ keV ($\sim$ 30\,000 km s$^{-1}$) and depth (measured with the {\tt gabs} model) $A\sim0.3-1.2$ keV (see Fig.~\ref{fig:outflow_spectra}). 
Such a dramatic change in line properties is unprecedented in both QPEs and TDEs, making an immediate explanation difficult. \cite{2025ApJ...984..124C} suggest that the evolution is consistent with homologous expansion of shock-heated debris ejected during an orbiter-disk collision, but this interpretation faces some strain due to the extreme mass and energy budgets that would be required. Detailed spectral modeling and radiation transfer simulations are required to better understand the observed behavior.

%Follow-up observations will reveal whether similar properties are seen in other QPEs.
%\newline

\subsubsection{Constraints from coordinated observations}
Coordinated multi-wavelength observations of QPEs have the potential to provide powerful additional constraints to QPE properties and models, though they remain rare because of the difficulty of coordinating strict simultaneity between ground- and space-based observatories.

At radio wavelengths, the most sensitive time-resolved observations exist for GSN 069 and RXJ1301, neither of which shows correlated radio variability with X-ray eruptions down to the $\sim$few per cent level \cite{2019Natur.573..381M, 2024A&A...692A..15G}. A comprehensive inventory of radio emission by \cite{2025PASA...42..130G} reported on measurements or upper limits for 12/13 sources, although not all of these resolve the eruption timescales. While some stochastic variability is present on hour- to year-long timescales, no evidence for correlated variability was found, effectively ruling out the possibility of a common emission mechanism for the observed radio variability and the QPEs. 

In the UV/optical bands, contemporaneous observations are obtained while observing with Swift (the XRT and UVOT instruments) as well as XMM-Newton (the EPIC and OM instruments). Unfortunately, the large PSFs of the optical telescopes ($\sim$7-10 arcsec) mean that nuclear emission is heavily diluted/contaminated by host galaxy light. As a result, these observations typically provide only weak limits on UV/optical counterparts (e.g. \cite{2019Natur.573..381M, 2021Natur.592..704A}). 

Nevertheless, low-amplitude UV variability correlated with (but delayed and smeared relative to) the X-ray eruptions has recently been reported in ZTF19acnskyy \cite{2026ApJ..1000L..57G}. 
Based on 5 cycles, the UV variability was found to lag the X-ray eruptions by $\sim$1 day; it does not have the same distinct rapid rise gradual decay evolution; and it is limited to $\sim$25\% in amplitude, a factor of 500 smaller than the X-ray eruptions. These properties indicate that this is not the direct counterpart of the X-ray emission mechanism.

If the correlated behavior is interpreted as a reverberation signal, the time delay implies that the UV emitting region is located at $\sim$1000 gravitational radii \cite{2026ApJ..1000L..57G}. This may help explain why other QPEs, some of which have been found to have compact accretion disks $\lesssim 1000 R_g$, do not show similar behavior.
Alternatively, the same authors suggest that the 1 day time delay can also be produced by accounting for the longer diffusion timescales for UV photons (compared to the X-rays) in expanding shock-heated material. This could be consistent expectations from star-disk collision models, although the positive period derivative seen in this source is very challenging to explain in this scenario \cite{2026ApJ..1000L..57G}.

Other sensitive time-resolved UV constraints come from HST observations of eRO-QPE2, obtained over 10 orbits in coordination with XMM-Newton \cite{2025ApJ...980L...1W}. Scheduling such campaigns is challenging due to the limited HST visibility and orbital constraints, and the lack of precise X-ray ephemerides. This will make similar campaigns very difficult for longer QPE recurrence times.
No correlated UV variability was detected during these joint observations, placing stringent upper limits on the UV counterparts and correlated variability to these flares; \cite{2025ApJ...980L...1W} reported an upper limit for the ratio of UV to X-ray luminosity of L$_{\rm FUV}$ / L$_{\rm X}$ $< 16-85$ during the QPEs (with the large uncertainty driven by the poorly constrained host extinction). 

Better (i.e. more sensitive and higher cadence) observational coverage is needed to unambiguously establish the incidence of multi-wavelength counterparts to the X-ray flares in most of the sample; the discovery of correlated variability in at least 1 source is a strong motivator for coordinated multi-wavelength observations of other sources.

\subsection{Quiescence}
While QPEs are defined by their dramatic flares, observational insight also comes from the properties of their quiescent emission: most QPE sources exhibit a persistent, lower-level emission component that is detectable outside of the eruptive phases. This component has been detected at X-ray, UV, optical and NIR bands. It provides a complementary probe of QPE systems, offering insight into the structure, size, and long-term evolution of the accretion disk independent of the eruption mechanism. \\

The quiescent emission is generally well-described by multi-color thermal disk models assuming local blackbody-like emission from each annulus, and an overall temperature gradient $T\propto T_{\rm in}(r/r_{\rm in})^{-3/4}$ (e.g. the \texttt{diskbb} model in \texttt{XSPEC}). The inferred inner disk temperatures are in the range $T_{\rm in}\sim50-100$ eV, typically a factor of a few smaller than the eruption temperatures. Crucially, the eruptions are not simply the high-luminosity tail of the quiescent emission component, indicated by the statistical preference of the peak spectra for an additional blackbody component on top of the disk emission over a single multicolor disk blackbody in multiple QPE sources \cite{2021ApJ...921L..40C,2024A&A...692A..15G}. This indicates that QPEs are instead a physically distinct emission process superimposed on an underlying accretion disk.

\subsubsection{Thermal vs. non-thermal components}
All QPEs are consistent with eruption peaks dominated by thermal blackbody-like emission and quiescent emission dominated by multicolor disk blackbodies. However, some QPEs show weak non-thermal power-law emission in quiescence, which is reminiscent of the corona in AGN and some TDEs \cite{2025ARA&A..63..379K}. This non-thermal emission is seen only during quiescence and at higher energies $\gtrsim 0.8$ keV, where the disk thermal emission becomes fainter. Such non-thermal emission has been detected in GSN 069, RX J1301, 2MASXJ0249, eRO-QPE3, and AT2022upj \cite{2019Natur.573..381M,2020A&A...636L...2G,2021ApJ...921L..40C,2024A&A...684A..64A,2025ApJ...983L..39C}. 

%In GSN 069, the power-law spectral component shows a luminosity of $(0.7-1.4)\times 10^{40}$ erg s$^{-1}$ (about $9\times$ fainter than the disk thermal emission), can be described by a power-law photon index $\Gamma=1.8$ (although it is poorly constrained), and is weak but detected at all epochs. In RX J1301, the power-law component increased by a factor of a few between 2000 and 2019, from $1$ to $3.9\times 10^{-14}$ erg s$^{-1}$, which is $7 (2)\times$ fainter than the disk brightness, respectively. In most epochs $\Gamma$ is unconstrained, but generally $1.9< \Gamma <3.7$ \cite{2024A&A...692A..15G}. 

%For 2MASXJ0249, the non-thermal component emerged several years after the initial TDE peak---after the QPEs had ceased---at a flux level $\sim 7\times$ fainter than the thermal disk emission, with $\Gamma=1.8\pm1.6$. 

%In eRO-QPE3, the power-law component was only detectable at early times (when the quiescent emission was brighter) at a consistent flux level of $(2.4-2.9)\times 10^{-13}$ erg s$^{-1}$ cm$^{-2}$, which is $4.8/1.3\times$ the disk thermal emission in the eRASS1/eRASS2 epochs, respectively \cite{2024A&A...684A..64A}. In subsequent epochs, the quiescent emission was too faint to be detectable. The non-thermal emission evolved similarly in AT2022upj, showing $2.8\times$ \textit{higher} bolometric luminosity ($7.4\times10^{42}$ erg s$^{-1}$) than the quiescence at early times ($1-2$ years post-optical peak), then eventually fading below detection limits at later times ($\gtrsim 2$ years) when QPEs emerged \cite{2025ApJ...983L..39C}.

This component shows a wide range of behaviors. In GSN 069 and RX J1301 the power-law component has remained roughly constant in flux and slope throughout the long-term evolution, even as the disk emission has varied considerably; in 2MASXJ0249 the non-thermal tail emerged most strongly at later times, after the QPEs disappeared; and in eRO-QPE3 and AT2022upj it was strongest at early times before eventually fading---along with the thermal disk emission---below detection thresholds. 

In the other 7/13 currently known QPE sources, limits on the presence of non-thermal emission are weaker, primarily due to the relative faintness of their quiescent emission, but the presence of a low-luminosity, power-law X-ray component cannot be ruled out. %Perhaps the observed diversity is not surprising given that both AGNs and TDEs are known to show a wide range of X-ray behaviors and corona evolution.

\subsubsection{QPEs as probes of accretion disk structure}
%The presence of a stable quiescent emission component enables a range of analyses that are not possible using the eruptions alone. In particular, modeling the broadband spectral energy distribution (SED) allows constraints to be placed on the radial extent, mass, and viscous evolution of the accretion disk, as well as on the mass of the central black hole. In the following subsections, we highlight how quiescent emission can be used as a powerful probe of accretion disk structure. 
The joint modeling of the X-ray--UV--optical SEDs of QPE sources has proven to be a powerful probe of accretion disk and SMBH properties by providing stringent constraints on the size of the accretion disk \cite{2025ApJ...980L...1W, 2025ApJ...985..146G}. Based on modeling employing a classical Shakura-Sunyaev thin disk framework with a null-stress boundary condition at the inner edge, and accounting for radiative transfer effects in the disk atmosphere \cite{2025ApJ...978..167G}, the quiescent UV--X-ray emission of every QPE for which data of sufficient quality is available to date is consistent with compact accretion disks with sizes (i.e. outer radii) ranging from a few 100 to a few 1000 R$_g$. This is incompatible with the typical sizes of AGN disks ($\gg 1000$s of R$_g$), but consistent with both expectations and observations of newly created disks following stellar disruption \cite{1988Natur.333..523R, 2025arXiv251026774G}. 

\begin{figure*}
    \centering
    \includegraphics[width=\textwidth]{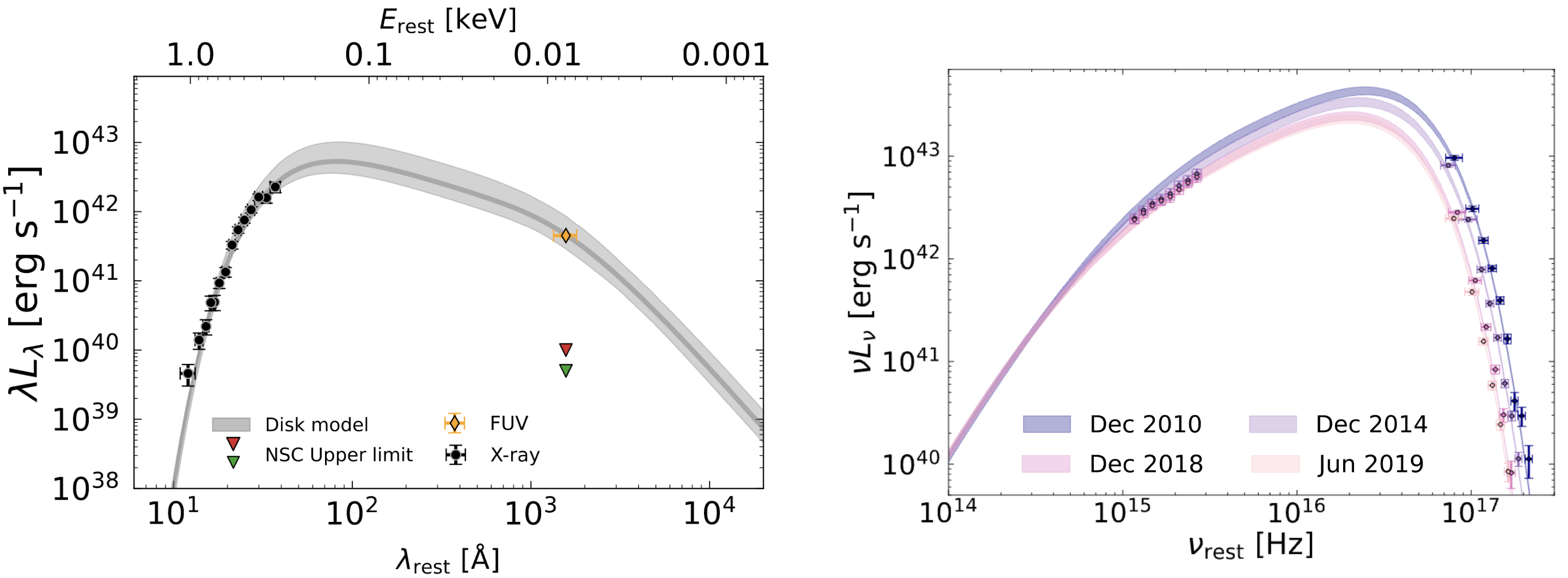}
    \caption{Spectral energy distributions (SEDs) of quiescent emission in eRO-QPE2 (\textbf{left}) and GSN 069 (\textbf{right}) fit with models of a compact accretion disk. X-ray data are from \textit{XMM-Newton} and UV data are from the Hubble Space Telescope. Figures reproduced from \cite{2025ApJ...980L...1W} and \cite{2025ApJ...992..114G}.}
    \label{fig:quiescence_SEDs}
\end{figure*}

The QPE sources following optically-discovered TDEs (e.g. AT2019qiz \cite{2024Natur.634..804N}, AT2022upj \cite{2025ApJ...983L..39C}) offer an even more powerful probe of the quiescent accretion disk, as they allow drawing upon the decades of detailed modeling and observations of TDEs (e.g. \cite{2021ARA&A..59...21G}) to study the quiescent emission from a multi-wavelength perspective. Multi-band photometry of the late-time plateau emission in TDEs is compatible with expectations for a compact disk containing $\sim 0.5M_\odot$ of material undergoing viscous expansion from the tidal radius \cite{2020MNRAS.492.5655M}, and photometric models have correspondingly been developed to use the late-time plateaus for parameter inference of the newborn disk and the SMBH (e.g. \cite{2025MNRAS.544.2225M}). 

\begin{figure*}
    \centering
    \includegraphics[width=\textwidth]{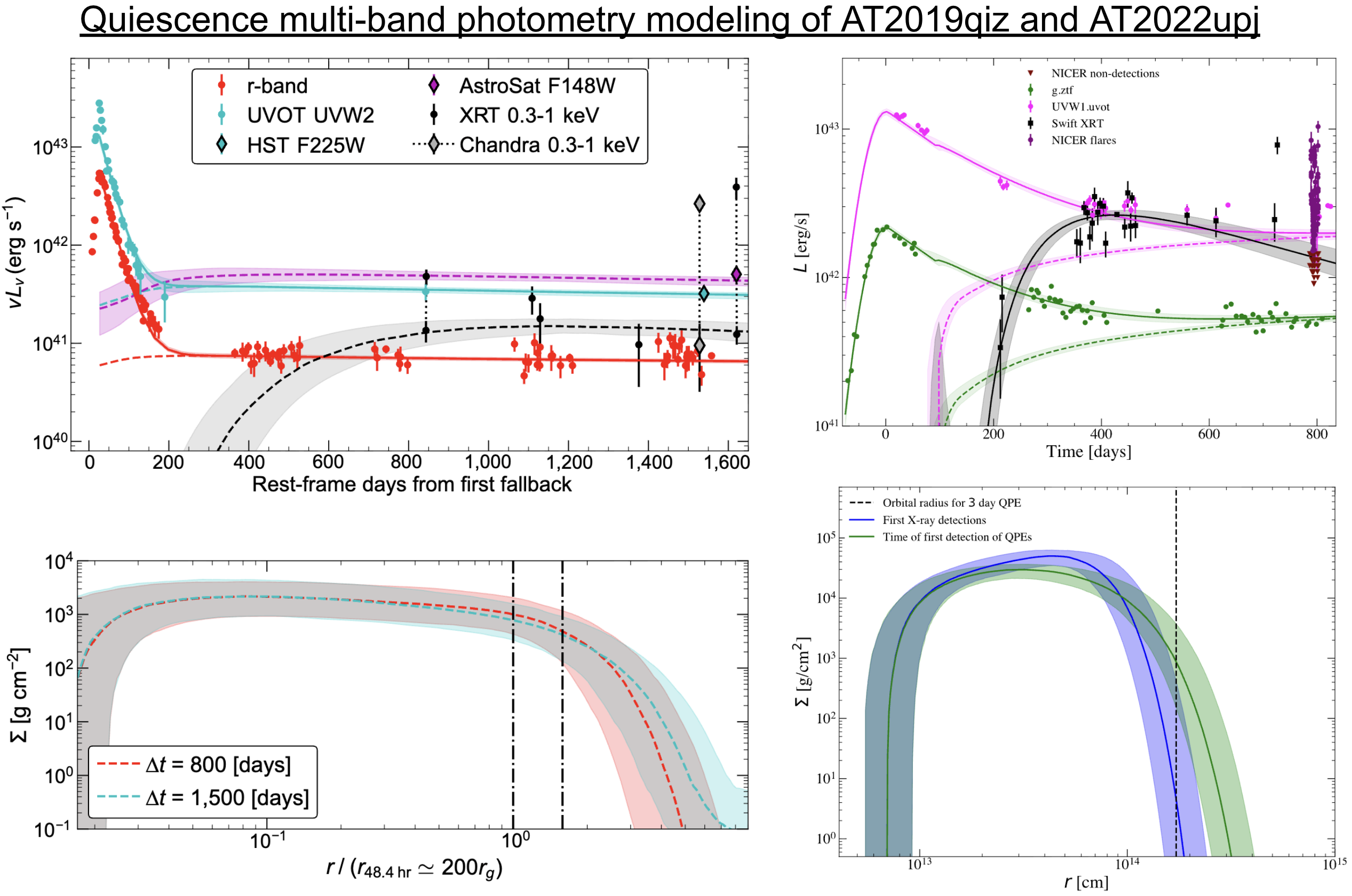}
    \caption{Multi-band photometry (optical/UV/X-ray) of early-time TDE flare and late-time plateau emission in TDE+QPE sources AT2019qiz (left) and AT2022upj (right). \textbf{Top panels:} Multi-band photometry of the TDE and onset of QPEs in both sources. \textbf{Bottom panels:} Model fits assuming a compact, viscously expanding accretion disks show the disks expand to the putative orbital radius by the onset of QPEs. Figures are reproduced from \cite{2024Natur.634..804N} and \cite{2025ApJ...983L..39C}.}
    \label{fig:quiescence_photometry}
\end{figure*}

Multi-band photometric fits have been performed to the late-time plateau emission of the TDEs AT2019qiz and AT2022upj to infer the disk surface density profiles and radial extents (\cite{2025MNRAS.544.2225M}; Figure \ref{fig:quiescence_photometry}). Both sources are consistent with compact accretion disks whose outer radii encompass the inferred orbital radius from the QPE recurrence time \cite{2024Natur.634..804N, 2025ApJ...983L..39C}, consistent with EMRI-disk collision models (see also \S \ref{sec:models}). The case of AT2022upj is especially suggestive, as the disk viscously expanded to the inferred radius over the course of $\sim$one year, before which no QPEs were observed. Then, the data after the onset of QPEs shows a disk whose outer edge roughly coincides with the relevant orbital radius \cite{2025ApJ...983L..39C}.

\subsubsection{Time-dependent modeling of viscously evolving disks}
Building on these constraints, recent work has extended quiescent disk modeling to a fully time-dependent and general relativistic framework, allowing the viscous evolution of the accretion flow to be traced across multiple epochs and providing tighter constraints on disk physics and timescales \cite{2025ApJ...992..114G}. By solving the thin disk equations for a particular SMBH mass, disk mass, and viscous timescale, one can compare predictions to multi-epoch data of the full disk SED to perform Bayesian parameter estimation and infer the disk parameters. These models have provided good fits to the available multi-epoch X-ray and UV data in GSN 069, providing the most informative constraints to date on the properties of the accretion disk emission underlying QPEs. 

One emerging result of these analyses is that QPE accretion disks have unusually long viscous timescales (e.g. $\gtrsim 2000$ days in GSN 069), which encapsulates uncertainties in the efficiency of mass/angular momentum transport due to magnetic stresses in the disk \cite{2025arXiv251026774G}. Though the origin of the variation in viscous timescales in accretion disks is unknown, it is possible that there is some link with the conditions necessary to trigger QPEs. \\

Taken together, the properties of the quiescent emission in QPE sources point to a physical picture in which the underlying accretion flow is compact, thermally dominated, and fundamentally different from the extended disks seen in typical AGNs. The inferred disk sizes of a few hundred to a few thousand gravitational radii are incompatible with standard AGN accretion flows, but are naturally explained by newly formed disks following stellar disruption. These findings provide strong empirical support for a close connection between QPEs and TDEs, and place important constraints on theoretical models for the origin of the eruptions. In particular, any viable scenario must account for the presence of a long-lived, compact accretion disk that persists independently of the QPEs, a requirement we revisit in detail when discussing proposed QPE formation mechanisms in \S \ref{sec:models}.

% \bibliographystyle{style/spphys.bst}
% \bibliography{references.bib}

% \section{Population Properties}
% \subfile{4_populationsection}

\setcounter{section}{3}
\section{Demographics}
With the growing sample size it is becoming possible (albeit still subject to significant uncertainties) to characterize the population properties of these systems. In this Section we describe efforts to constrain the formation rates of QPEs and implications for their nature, after which we highlight the heterogeneity of the disparate selection functions of the channels through which QPEs have been discovered, and the potential impact on the observed properties.

\subsection{Rates}
\label{sec:rates}
QPEs are a rare phenomenon, with only 13 known sources to date. However, quantifying this more precisely, i.e. converting the size of the observed sample into an intrinsic rate, is a complex endeavor. The most significant unknowns in this conversion are sample completeness and the effect of selection biases. The heterogeneous nature and selection of the current sample makes estimating the completeness, and therefore the QPE rate, difficult. There are two main methods that have been applied to try to estimate the intrinsic QPE rate: one based solely on blind survey discoveries (the eROSITA-selected QPEs), and one based on targeted searches (QPEs detected at late times in TDEs).

\begin{figure}
    \centering
    \includegraphics[width=\linewidth]{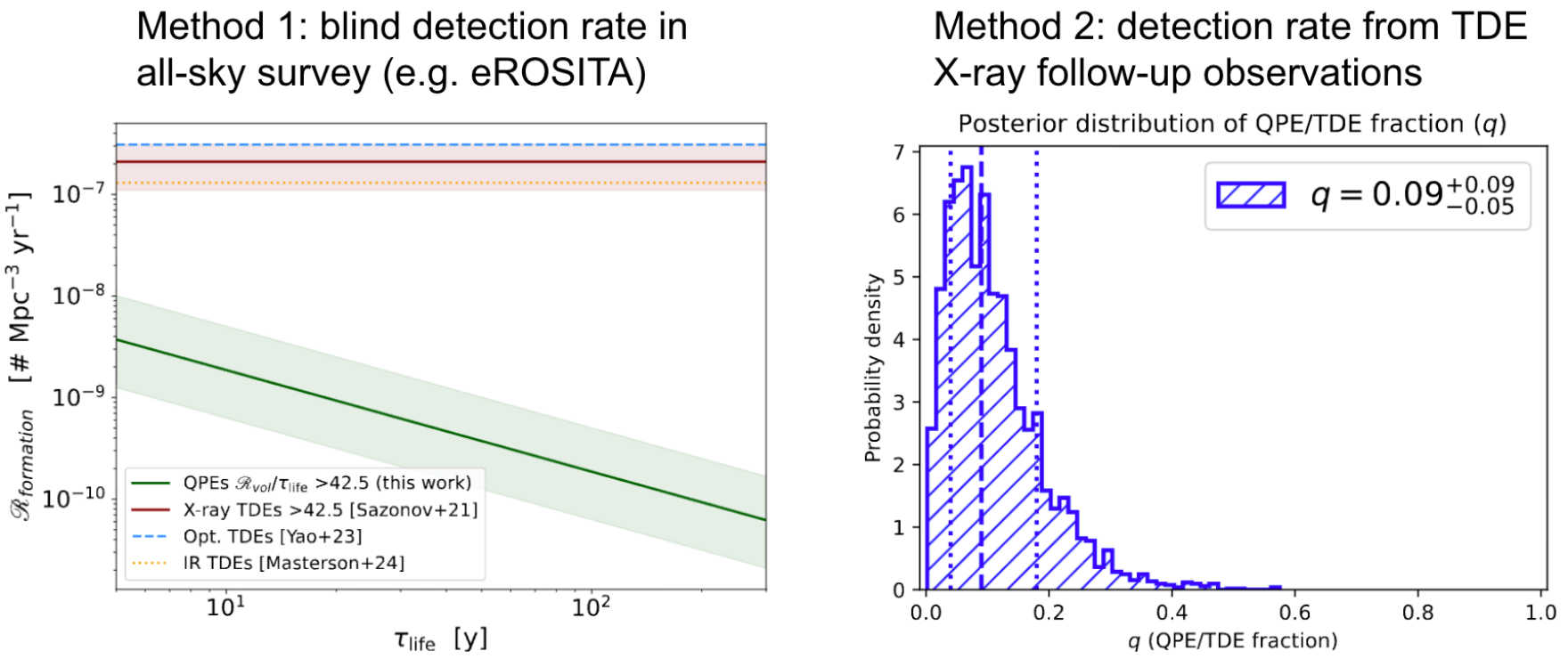}
    \caption{Rate estimates from the two methods discussed in Section
    \ref{sec:rates}. \textbf{Method 1} \cite{2024A&A...684L..14A} relies on the sensitivity of the eROSITA all-sky survey to obtain an unbiased estimate of the detection rate. \textbf{Method 2} \cite{2025ApJ...983L..39C} relies on the QPE-TDE association, and quantifies the likelihood of detecting a population of QPEs after a fixed number of X-ray follow-up observations of optical TDEs.}
    \label{fig:rates}
\end{figure}

\subsubsection{Rate estimates from a blind survey}
In order to minimize the effects of unknown selection biases in the QPE sample, one can focus on a sub-sample with clear selection criteria, such as that provided by the eROSITA-selected sample \cite{2024A&A...684L..14A}. As this observatory performed repeated scans of the sky, the blind QPE search provided a relatively homogeneous sample, where the main selection effects (e.g. luminosity and recurrence time) can be quantified in a straight-forward manner. This method provides an instantaneous volumetric rate ("How many QPEs are active at a given moment in a given volume of the Universe?"), and can be converted into a formation rate ("How many QPEs are created in a given volume of the Universe in a given period of time?"). It is the latter quantity that can ultimately provide more insight into viable QPE formation channels. 

The eROSITA-based search was limited to QPEs with a peak luminosity brighter than log($L_{\rm 0.5-2.0~keV}^{\rm peak})>41.7$ erg s$^{-1}$. The instantaneous volumetric rate of QPEs estimated through this method is $\mathcal{R}_{\rm vol}~=~0.60^{+4.73}_{-0.43}\times 10^{-6}$~Mpc$^{-3}$, corresponding to a per galaxy rate of $\mathcal{R}_{\rm gal}~=~0.36^{+2.87}_{-0.26}\times 10^{-4}$~gal$^{-1}$ (Fig.~\ref{fig:rates}, left panel). To convert this instantaneous rate into a formation rate, one must take into account the temporal evolution of QPEs, and their lifetime. The simplest assumption is that QPEs are active and detectable over a typical timescale $\tau_{\rm life}$, which is poorly constrained at present. 
% The formation rate is calculated by dividing the instantaneous rate by this lifetime. 
This yields a volumetric formation rate of $\mathcal{R}_{\rm vol}/\tau_{\rm life}\sim0.6\times 10^{-7}~(\tau_{\rm life}/10~{\rm yr})^{-1}$~Mpc$^{-3}$~yr$^{-1}$, and a formation rate per galaxy of $\mathcal{R}_{\rm gal}/\tau_{\rm life}\sim0.4\times10^{-5}
~(\tau_{\rm life}/10~{\rm yr})^{-1}$~gal$^{-1}$~yr$^{-1}$. 

The latter quantity can be compared directly to the occurrence rate of TDEs to provide insights into their tentative connection. TDE rates have been estimated from various wavelength regimes and surveys to be of order $\mathcal{R}_{\rm TDE}\sim10^{-5}-10^{-4}$~gal$^{-1}$~yr$^{-1}$, depending on the wavelength and type of the host \citep[e.g.][]{2020SSRv..216...32F, 2023ApJ...955L...6Y, 2025A&A...697A.159G}. This would suggest that, if \textit{all} QPEs are indeed physically linked to TDEs, then a fraction of order $q$ = $\mathcal{O}(0.1)$ of TDEs should ultimately lead to the creation of a QPE source.

\subsubsection{Rate estimates from QPEs associated with TDEs}
Another approach to quantify the QPE rate is to exploit the coincidence between QPEs and TDEs by focusing on X-ray observations taken at late times in TDEs \cite{2025ApJ...983L..39C}. This method requires quantifying the non-unity detectability of QPEs, which can be active but remain undetected due to variations in their timing properties and/or luminosity (as discussed in Sec. \ref{sec:eruptions}). 

For a given observing strategy (number of snapshots, total duration of exposures) the detection or non-detection of QPEs contributes some information about the overall QPE/TDE occurrence fraction, $q$, assuming it is a constant for all TDEs. One can  assemble a population-wide likelihood function to perform Bayesian inference using a seed population of TDEs and a set of X-ray observations for each source. \cite{2025ApJ...983L..39C} assembled an exhaustive inventory of the X-ray coverage of all optically selected TDEs from the Transient Name Server\footnote{https://www.wis-tns.org}, amounting to 100 TDEs, in which three showed QPEs -- AT2019qiz, AT2019vcb, and AT2022upj. Estimating $q$ via Bayesian nested sampling requires specifying priors on several QPE properties, which we summarize here: 
\begin{itemize}
    \item Lifetime, drawn from a uniform prior distribution $\mathcal{U}(1,20\,{\rm yr})$.
    \item Recurrence time, drawn from a uniform prior distribution $\mathcal{U}(0.1,5\,{\rm days})$.
    \item Time delay between disruption and the onset of QPEs, drawn from a uniform prior distribution $\mathcal{U}(0.1,5\,{\rm yr})$.
    \item QPE peak luminosity, drawn from a power-law prior distribution following $P({\rm log}( L_{\rm peak}))\propto -1.83\times{\rm log} (L_{\rm peak})$.
\end{itemize}

While some of these assumptions may need to be revised in light of discoveries that expand the observed QPE parameter space, the results are not expected to be strongly sensitive to modest changes in the prior bounds. The resulting estimate of the fraction of TDEs hosting QPEs within 5 years of disruption is $q=0.09^{+0.09}_{-0.05}$ (Fig.~\ref{fig:rates}, right panel). This is in remarkable agreement with the instantaneous rate per galaxy as derived from the eROSITA sample, and leads to a rate per galaxy per year $\mathcal{O}(0.1)$ of the TDE rate. Note that this method only applies to QPEs discovered following optical TDEs, and it remains unclear whether that is the only formation scenario.

With these rates estimates in hand, one can assess consistency with various QPE models. For instance, in the EMRI model, assuming that the initial EMRI and the secondary TDE that creates the accretion disk are independent, a rate of QPEs per TDE of $q=0.09^{+0.09}_{-0.05}$ suggests that, even in the absence of TDEs to reveal the EMRI in EM radiation, almost 10\% of TDE host galaxies should host an EMRI at $\mathcal{O}(100R_g)$ of their central SMBH at any given time. This high implied fraction of (potentially) gravitational wave-emitting sources could have significant implications for the future LISA mission, particularly in shaping expectations for the astrophysical background signal (see also \S \ref{sec:multimessenger}).

\subsection{Selection biases}
\label{sec:biases}
Interpreting the emerging population-level properties of QPEs requires careful consideration of the observational biases that shape the current sample. Given the small number of known sources and the heterogeneous discovery methods, apparent trends in timing, luminosity, or host-galaxy properties may reflect selection effects rather than intrinsic characteristics. 

% We outline the dominant biases that affect QPE detection and classification first, and then discuss how they can influence our current understanding of the QPE population.
% \begin{enumerate}
%     \item QPEs are easier to detect at intermediate duty cycles (10–50\%), where eruptions are frequent enough to increase the detection likelihood but isolated enough to be identifiable as individual flares.
%     \item Without well-defined criteria for what constitutes a QPE, sources that do not closely resemble known cases may be excluded, even if they share the same physical origin.
%     \item Due to their soft X-ray spectra, emission from QPEs at higher redshift will shift out of the observable band of current X-ray telescopes.
%     \item Recent QPE discoveries result from targeted late-time X-ray monitoring of known TDEs, skewing the sample toward TDE-associated systems.
% \end{enumerate}
% We describe each of these biases and their effects on the population properties in more detail below.

\subsubsection{Discovery bias in duty cycle parameter space}
%The defining observational signature of QPEs -- narrow, recurrent eruptions -- naturally introduces strong selection effects in timing parameter space. In practice, QPEs are most readily identified when the eruption duration and recurrence time conspire to produce a moderate duty cycle, such that individual flares are both frequent enough to be detected and sufficiently isolated to be recognized as discrete events.
\begin{figure}
    \centering
    \includegraphics[width=0.7\linewidth]{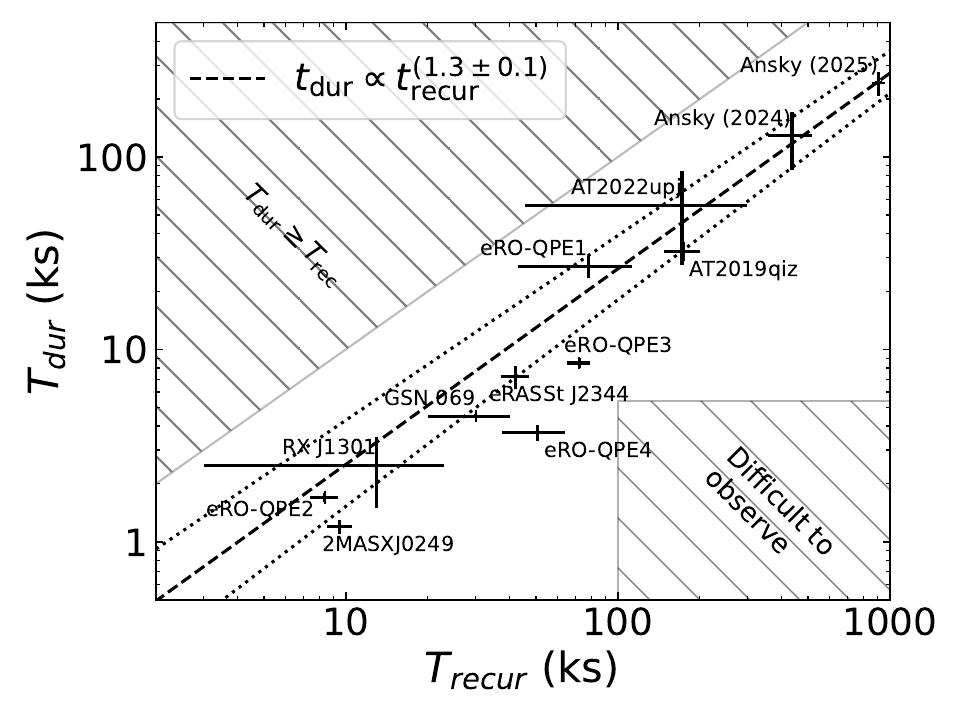}
    \caption{Timing properties of the current QPE sample, used as an illustration of the selection effects at play. Reproduced from \cite{2025A&A...703A.263H}.}
    \label{fig:SelBias}
\end{figure}

Fig. \ref{fig:SelBias} displays the timing properties of the current sample by comparing the peak duration $t_{\rm dur}$ to the recurrence time $t_{\rm recur}$. Most objects lie along a diagonal line, hinting at an intrinsic scaling between these two parameters. More quantitatively, the relation follows a power-law function of the form $t_{\rm dur}\propto t_{\rm recur}^{(1.3\pm0.1)}$ \cite{2025A&A...703A.263H}. However, this reveals the selection biases at play in our sample: sources away from the diagonal are harder to detect, and as such the apparent scaling may arise in part or in full from these observational selection effects. 

\begin{figure}
    \centering
\includegraphics[width=0.7\linewidth]{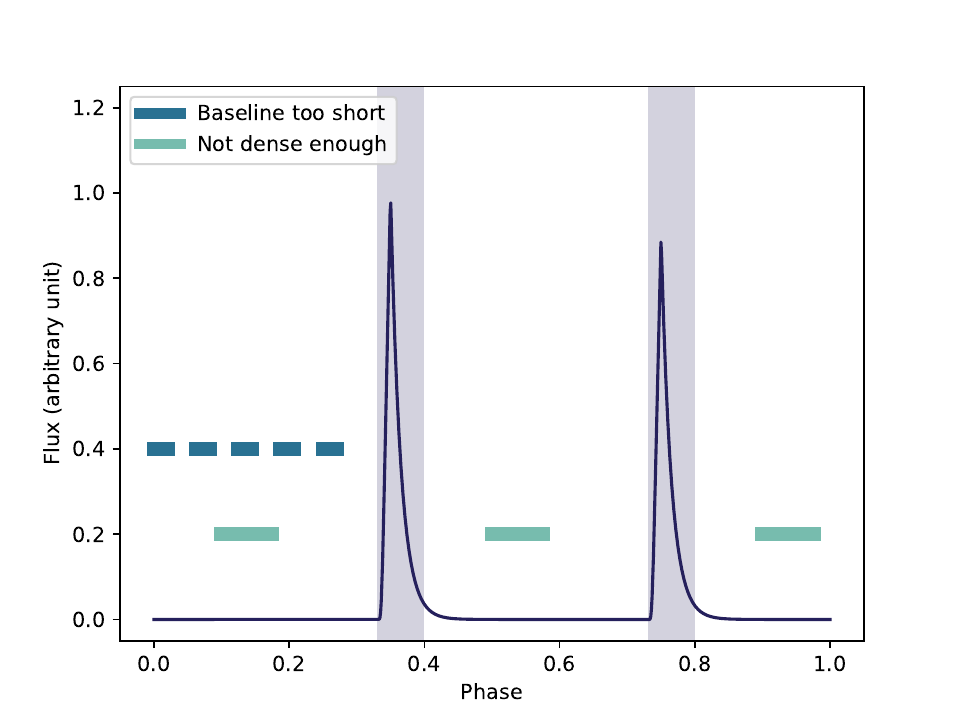}
    \caption{Illustration of the trade off between density and baseline length, given a constant exposure time. Dense monitoring will lead to a short baseline (blue lines), while a long baseline leads to sparser density (teal lines). Both scenarios have a significantly non-zero probability of missing the eruptions (shaded).}
    \label{fig:dutycycles}
\end{figure}

For points below the diagonal, the duration of the peaks is small compared to the recurrence time, and as such the duty cycle is low. This means that one needs to have access to high cadence temporal monitoring over a baseline longer than t$_{\rm recur}$ in order to detect eruptions, and thus these objects are likely under-represented in the sample. This effect is illustrated in Figure \ref{fig:dutycycles}. 

For points above the diagonal, the duration of each peak is larger than the recurrence time, and as such successive peaks overlap. This means that the lightcurve does not display the typical expected shape of QPEs and does not adhere to the QPE definition outlined $\S$ \ref{sec:properties}. One could imagine that peculiar variability in X-ray selected TDEs are the result of overlapping QPEs, i.e. the individual flares are not discernible. One potential instance of such an object is 2XMM J123103.2+110648 \cite{2025A&A...700A..48C}, whose light curve is reminiscent of quasi-periodic oscillations (QPOs) rather than QPEs, but whose X-ray spectral properties are in line with the current QPE population. 

Quantifying the intrinsic distribution of QPE timing properties therefore requires accounting for the strong suppression of sources with low and high duty cycles. This is challenging because it would require a careful estimate of the X-ray (inherently multi-mission) coverage of all potential QPE sources. If this selection effect is strong enough, then the inferred scaling between $t_{\rm recur}$ and $t_{\rm dur}$ may appear due to selection biases, and not correspond to an actual physical process. 

\subsubsection{Classification bias due to lack of well-defined criteria}
A second important bias arises from the absence of a formal, quantitative definition of what constitutes a QPE. As a result sources are classified as QPEs only if they closely resemble previously confirmed systems in recurrence, spectral softness, and hysteresis ($\S$\ref{sec:properties}).

%In practice, sources are typically classified as QPEs only if they exhibit multiple observational hallmarks closely resembling those of previously confirmed systems, including recurrence, spectral softness, and characteristic hysteresis patterns (see the working definition in presented in \S\ref{sec:properties}).

This conservative approach  guards against potential contaminants such as short-term flares in X-ray TDEs, but excludes candidates lacking sufficient temporal coverage or those observed during unfavorable phases (e.g. had eRO-QPE1 only been observed during its overlapping-burst phase, it may not have met the criteria for classification). Several strong QPE candidates remain unconfirmed due to these limitations; two examples are AT2019ehz \cite{2021ApJ...908....4V} and AT2019aalc \cite{2026A&A...706A.324V}. Similarly, repeating nuclear transients on different timescales (e.g. Swift J0230, \cite{2023NatAs...7.1368E, 2024NatAs...8..347G} and ASASSN--14ko, \cite{2021ApJ...910..125P}) and with somewhat different properties are not generally recognized as QPEs.
% \textcolor{purple}{(Add other examples if available.)}

%\subsubsection{Redshift bias}
%The predominantly soft X-ray spectra of QPEs impose a strong redshift-dependent selection bias. Because a significant fraction of the emitted energy lies outside of the low energy sensitivity threshold of current X-ray observatories, QPEs become increasingly difficult to detect with increasing redshift, even at fixed intrinsic luminosity.

%To date QPEs have been found in the nuclei of galaxies with redshifts ranging from $0.015 < z < 0.115$. The detection volume for detecting QPEs with peak luminosities up to 10$^{44}$ erg s$^{-1}$ in principle extends to z$\sim$1 (assuming a 1 ks XMM-Newton limiting depth of 2$\times$10$^{-14}$ erg cm$^{-2}$ s$^{-1}$ or a 3 ks Swift/XRT limiting depth of $\sim$10$^{-13}$ erg cm$^{-2}$ s$^{-1}$).
%However, as a consequence of their spectral properties, most of the photons are emitted in the far UV and the soft X-rays, typically up to $\sim$ 1.5 keV. 

%Existing X-ray observatories are sensitive down to 0.25 keV, and a significant fraction of the QPE photons are below the sensitivity cut-off of their instruments. As a direct consequence, it is very challenging to observe high redshift ($z>1$) QPEs in practice. With a temperature of 100 eV and neglecting the effects of absorption, $\sim20\%$ of the energy flux of the emission is below 0.2 keV when the source is at $z=0$. This fraction rises to 60\% at $z=1$, and 85\% at $z=2$. To detect sources at significant redshift, far UV observations may offer a more promising avenue for discovery. 

\subsubsection{Confirmation bias from TDE-triggered searches}
Finally, the recent success of dedicated late-time X-ray follow-up of TDEs \cite{2024Natur.634..804N, 2025ApJ...983L..39C, 2025NatAs...9..895H} introduces a confirmation bias of its own: if searches are predominantly conducted in TDE hosts, the sample will increasingly be dominated by TDE-associated systems regardless of whether all QPEs share that origin. This effect can be mitigated by continued blind and archival searches \cite{2025arXiv251122520Q}. \\

As a result of these biases, the known population is a highly filtered subset of the underlying phenomenon, which must be kept in mind when interpreting population trends or confronting models with observations. Expanding the sample through complementary discovery strategies and future instrumentation will be essential for disentangling intrinsic properties from observational biases.

%Finally, we highlight a bias linked to the most recent method to detect new QPE sources -- dedicated late-time X-ray follow-up of optical TDEs \cite{2024Natur.634..804N, 2025ApJ...983L..39C, 2025NatAs...9..895H}. While this method is efficient in increasing the QPE sample, it introduces a strong selection and confirmation bias. If QPE searches are predominantly performed at late times following TDEs, the QPE sample will over time become dominated by these objects, despite the fact that it is not known \textit{a priori} whether all QPEs are necessarily associated with TDEs. To mitigate this effect, it is important to continue pursuing QPE searches in alternate ways, including blind and archival searches \cite{2025arXiv251122520Q}. \\

%Taken together, these selection effects highlight that the known QPE population likely represents a highly filtered subset of the underlying physical phenomenon. Biases related to duty cycle, classification criteria, spectral softness, and discovery strategy all act to disfavor broad regions of parameter space, potentially obscuring the true diversity of QPE behavior. These limitations must be kept in mind when interpreting population trends, comparing QPEs to other classes of nuclear transients, and confronting theoretical models with observations. Expanding the sample through complementary discovery strategies and future instrumentation will be essential for disentangling intrinsic properties from observational biases.

\section{Host Galaxies and Environments}

\setcounter{section}{5}
\label{sec:host}
% Galaxy types, redshift distribution.\newline
% Nuclear activity and SMBH mass estimates.\newline
% Stellar populations, morphology, environment.

The host galaxies of transients often provide complementary insights into their nature. For QPEs, host galaxy studies are particularly informative as they probe the black hole demographics, stellar population properties, and circumnuclear environments that may influence their formation and/or detection rate. 
%In this section we describe the QPE host galaxy properties based on available imaging, long-slit and spatially resolved spectroscopy data, including both published literature and results as well as several unpublished datasets. We also place QPEs in the broader context of tidal disruption events and low-mass black hole demographics. 
Two caveats should be kept in mind when interpreting the host analysis: i) small sample statistics, and ii) the fact that the analysis typically does not account for the presence of an accretion disk, which may influence the inferred properties (see e.g. \cite{2025arXiv251018985N} for a comparative analysis with/without a disk component). The properties of the host galaxies, discussed in the subsequent sections, are compiled in Table \ref{tab:host}.

%\subsection{Concentration and optical/IR colors}
%\subsubsection{Morphology and environment}
%The host galaxy morphologies of QPE sources include elliptical, S0 and spiral galaxies, which generally have smooth appearances in continuum light (see e.g. \cite{2025ApJ...994..209G}); only 1 source shows evidence for dust lanes in HST imaging (RXJ1301; \cite{1999AJ....118.1230C}). This suggests that QPE galaxies generally did not experience a major merger in the recent past (but see \S \ref{sec:eelr} for more details on merger history). 

%In terms of extragalactic environment, we perform a cross-match with cluster and galaxy group catalogues \cite{1989ApJS...70....1A, 2017A&A...602A.100T}, and find that 2/13 sources (RXJ1301 and ZTF19acnskyy) are associated with small galaxy groups of 3-4 members \cite{2000MNRAS.318..309D, 2017A&A...602A.100T}. This is consistent with the properties of the general galaxy population (i.e. QPE hosts do not prefer different environments), similar to TDE host galaxies \cite{2014A&A...566A...1T, 2020SSRv..216...32F}.

%Beyond their broad morphological classifications, QPE host galaxies do exhibit several structural properties that differentiate them from the general galaxy population.

\subsection{Concentration}
\cite{2025ApJ...994..209G} performed a homogeneous analysis of deep continuum images through bulge-disk decomposition, showing that QPE hosts are centrally concentrated (Sérsic indices n $\sim$ 3), bulge-dominated ($g$-band bulge-to-total light ratio B/T $\sim$ 0.5), and have high surface mass densities ($\sim$10$^{9-10}$ M$_{\odot}$ kpc$^{-2}$). These values are significantly higher than found in a mass- and redshift-matched control sample of SDSS galaxies. QPE hosts share these morphological properties with TDE hosts \cite{2017ApJ...850...22L, 2018ApJ...853...39G, 2020SSRv..216...32F,2024arXiv240910486G}. We have performed the analysis described in \cite{2025ApJ...994..209G} for 3 additional sources, providing a complete census of QPE host galaxy properties (see Table \ref{tab:host}). The only exception is the source eRASSt J2344, which is blended with nearby sources in the Legacy Survey imaging. %The Sérsic index for this source tends towards the upper allowed bound of $n = 9$ during fitting; higher spatial resolution imaging is required for a robust analysis.

%{\bf These structural signatures are typical of galaxies that have recently undergone an interaction or merger, processes that quench star formation and increase stellar densities in galactic nuclei. Such conditions raise the likelihood of stars scattering into the loss cone of the central black hole, potentially increasing the TDE rate and/or the QPE rate.}
%These characteristics are also closely tied to the stellar populations of the host galaxies, which can be probed through their optical and infrared colors.

\subsection{Optical and IR host colors}
While the overrepresentation of TDE host galaxies in the green valley, where galaxies are rapidly transitioning from the star-forming blue cloud to the red sequence, has been studied in detail (e.g. \cite{2021ApJ...908L..20H}), this has not yet been investigated for QPE hosts. We compile the $u-r$ color for all sources in Table \ref{tab:host}. For AT2019qiz and AT2019vcb, we take the values provided by \cite{2023ApJ...955L...6Y}. For the eROSITA QPEs, we compile archival UV, optical and NIR data and follow the SED fitting procedure outlined in \cite{2023ApJ...955L...6Y} to compute (rest-frame) synthetic $u-r$ colors.

\begin{figure}
    \centering
    \includegraphics[width=0.7\linewidth]{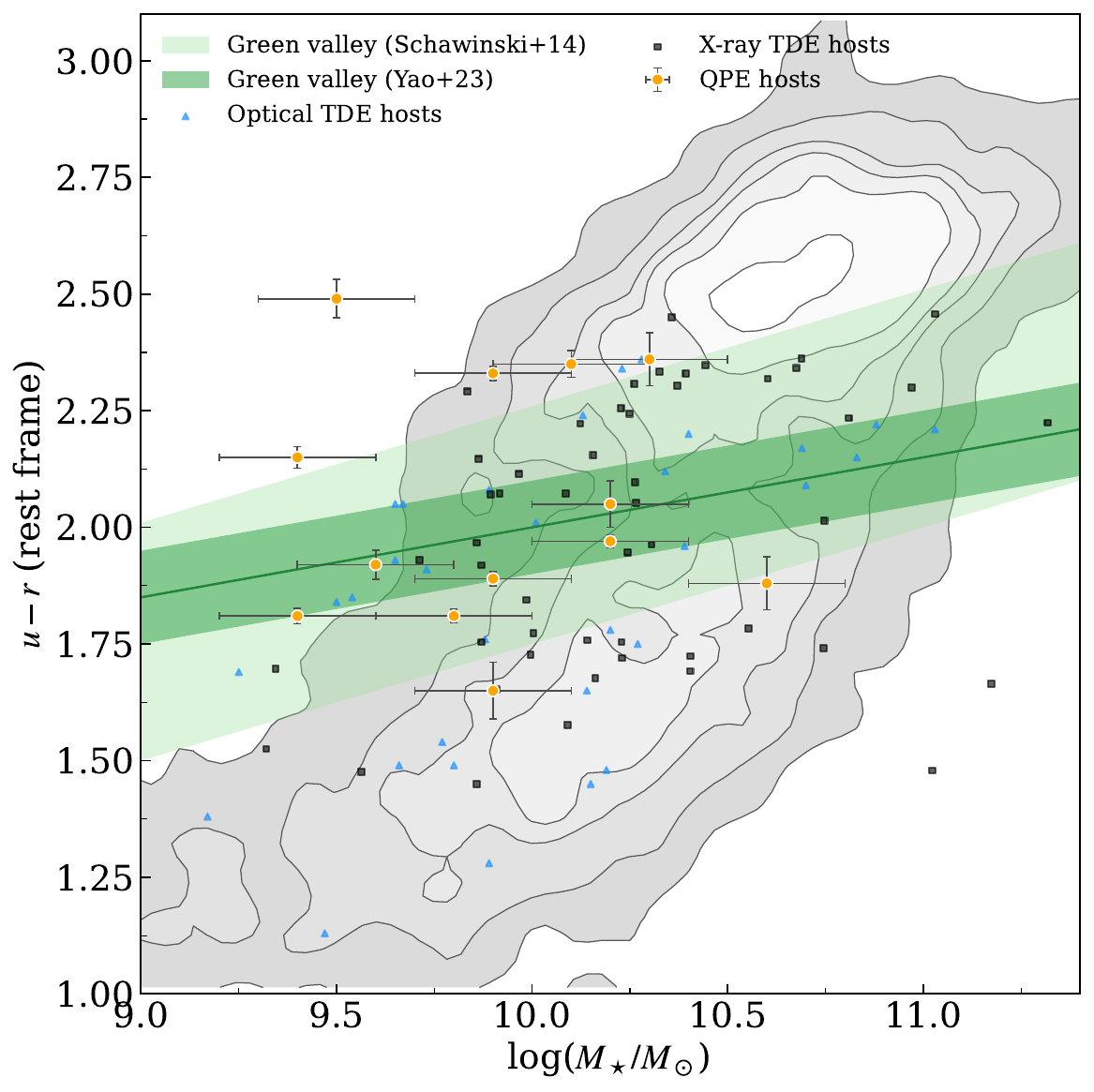}
    \caption{{\bf QPE hosts and the green valley.} Orange markers indicate QPE hosts, which are overlaid on an SDSS galaxy sample (grey contours). The light green band shows the green valley as defined in \cite{2014MNRAS.440..889S}, while the dark green band shows the definition of \cite{2023ApJ...955L...6Y}. Blue triangles and black squares show optical and X-ray selected TDE hosts from \cite{2023ApJ...955L...6Y} and \cite{2025arXiv251212480Z} for comparison. }
    \label{fig:greenvalley}
\end{figure}

Combined with the colors and stellar mass estimates from \cite{2025ApJ...994..209G}, this allows an assessment of how many sources are located within the green valley, shown in Figure \ref{fig:greenvalley}. Taking the general definition of \cite{2014MNRAS.440..889S}, we find that 6/13 (46 per cent, or within the range 0.29--0.65 accounting for Poisson statistics at 68\% confidence interval \cite{1986ApJ...303..336G}) sources fall within this region. A more restrictive definition of the green valley used by \cite{2023ApJ...955L...6Y} for a TDE host analysis yields similar results (0.39$\pm$0.16 within the green valley). This is significantly higher than the 12$\%$ of all SDSS galaxies within this region, demonstrating an overrepresentation of QPE hosts in the green valley\footnote{K-corrections to the observed optical colors in the relevant redshift range are small ($<0.05$) and do not change this conclusion.}. 

These results are remarkably consistent with the overrepresentation seen in TDE host galaxies ($\sim$50--60 per cent, \cite{2021ApJ...908L..20H, 2023ApJ...955L...6Y}). This effect is not driven by the inclusion of QPEs selected through optical TDE follow-up, since those three hosts fall outside of the green valley definitions.

%The implication is that QPE hosts preferentially reside in galaxies undergoing rapid evolutionary transitions, where changes in their central stellar and gas distributions can potentially enhance the rate of dynamical interactions in the nucleus.

Finally, the IR color (WISE W1-W2) can be used to assess the possible presence of an AGN and dusty torus; typical AGN selection criteria require W1-W2$>0.8$ \cite{2012ApJ...753...30S}. The QPE nuclei are relatively blue (W1--W2$<$0.3), indicating that none of the QPE hosts conform to the AGN selection and that their nuclei do not contain significant amounts of dust. This is consistent with the lack of cold absorption in the X-ray spectra.

% morph all kinds, B/T, sersic index, surface mass density, green valley overrepresentation, wise colors 

% table setup
\setlength{\tabcolsep}{3pt}
\renewcommand{\arraystretch}{1.2}
\begin{sidewaystable}[]
    \centering
    \begin{tabular}{c|cccccccccccccc}
Source &  z & (B/T)$_g$ & Sersic & $\Sigma_{M_{\star}}$ & $\log (M_{\star}$) & $u-r$ & W1-W2 & BPT & EELR & $\sigma_{\star}$ & M$_{\rm BH}$ & Green \\
& & & & (M$_{\odot}$/kpc$^2$) & (M$_{\odot}$) & (mag) & (mag) & &  & (km s$^{-1}$) & (log$_{10}$[M$_{\odot}$]) & valley \\\hline

AT2019vcb   & 0.088 & $0.53\pm$0.03 & $2.03^{+0.09}_{-0.03}$ & $10.01\pm$0.2 & $9.9\pm$0.2  & $1.65 \pm 0.06$ & 0.25$\pm$0.12 & SF & N &  --- & --- &  N \\
RXJ1301  & 0.024 &  $0.17\pm$0.01 & $2.67\pm$0.01 & $9.56\pm$0.2  & $10.2\pm$0.2 & $1.97 \pm 0.01$ & 0.02$\pm$0.03 & Non-stellar & Y   &  87(2) & 6.67$\pm$0.05$\pm$0.55 & Y\\
ZTF19acnskyy &      0.024  &  0.41$\pm$0.02 &    1.76$\pm$0.02&  9.66$\pm$0.2   & $10.2\pm$0.2 & $2.35 \pm 0.03$ & -0.03$\pm$0.05 & Non-stellar & Y & 75(3) & 6.37$\pm$0.08$\pm$0.55 & N \\
% 0.487 kpc / "; single sersic fit gives Re = 2.65+-0.01, coresponding to 1.29+-0.01 kpc
2MASXJ0249  & 0.019 &  $0.79\pm$0.01 & $2.20\pm$0.01 & $8.33\pm$0.2  & $10.3\pm$0.2  & $1.92 \pm 0.03$ & 0.05$\pm$0.04 & SF/Non-stellar & Y &  43(3) & 5.27$\pm$0.14$\pm$0.55 & Y \\
eRO-QPE4   & 0.044 &  $0.57\pm$0.02 & $3.26\pm$0.02 & $9.80\pm$0.2  & $10.2\pm$0.2 & 2.05 $\pm$ 0.05 & 0.07$\pm$0.04 & Non-stellar & Y   &  38(5) & 5.03$\pm$0.26$\pm$0.55 & Y \\
AT2019qiz & 0.015 &  $0.54\pm$0.01 & $3.74\pm$0.01 & $8.33\pm$0.2  & $10.3\pm$0.2 & $2.36 \pm 0.06$ & -0.04$\pm$0.03 & Non-stellar & Y &  70(2) & 6.24$\pm$0.06$\pm$0.55 & N \\
eRO-QPE1  & 0.050 &  $0.36\pm$0.02 & $2.02\pm$0.02 & $9.61\pm$0.2  & $9.9\pm$0.2  & 2.33 $\pm$ 0.02 & 0.06$\pm$0.09 & Non-stellar & N &  62(3) & 6.00$\pm$0.10$\pm$0.55 & N \\
AT2022upj &      0.054  & 0.25$\pm$0.02     &    6.00$\pm$0.08         &          8.78$\pm$0.2            & $9.5\pm$0.2  & $2.49 \pm 0.04$ & 0.06$\pm$0.04 &  SF & N  & --- & --- & N  \\
% halflight radius of AT2022upj with Gilbert's fitting code is 2.21" (1.057 kpc / ") = 2.34+-0.01 kpc. gives a stellar mass density of 8.78+-0.2
eRO-QPE3  & 0.024 &  $0.97\pm$0.02 & $2.80\pm$0.04 & $9.84\pm$0.2  & $9.4\pm$0.2  & 2.15 $\pm$ 0.02 & -0.32$\pm$0.10 & SF & N   & 21(7) & 3.86$\pm$0.66$\pm$0.55 & N \\
GSN069  & 0.018 &  $0.38\pm$0.01 & $5.06\pm$0.01 & $8.91\pm$0.2  & $9.8\pm$0.2  & $1.81 \pm 0.01$ & 0.01$\pm$0.03 & Non-stellar & Y   &   64(4) & 6.06$\pm$0.12$\pm$0.55 & Y \\
eRO-QPE2 & 0.018 &  $0.39\pm$0.01 & $1.18\pm$0.01 & $9.35\pm$0.2  & $9.4\pm$0.2  & 1.81 $\pm$ 0.02 & 0.16$\pm$0.04 & Non-stellar & Y   &  37(3) & 4.98$\pm$0.16$\pm$0.55 & Y  \\
eRO-QPE5 & 0.116 & 0.10$\pm$0.01 & 1.94$\pm$0.01 & 8.43$\pm$0.2 & 9.9$\pm$0.2 & 1.89 $\pm$ 0.03 & 0.32$\pm$0.05 & SF/Non-stellar & Y & 60(6) & 5.93$\pm$0.20$\pm$0.55 & Y  \\
% qpe5 - pixel scale is 2.099 kpc / ",  halflight radius (single sersic fit) of 2.6+-0.02" gives value of 5.46+-0.04 kpc
eRASSt J2344 & 0.10 & --- & --- & ---  & 10.6$\pm$0.2 & 1.88$\pm$0.04 & 0.38$\pm$0.04 & Non-stellar & Y &  106(10) & 7.06$\pm$0.19$\pm$0.55 & N \\
% erasj2344 sersic want to go to 9, likely due to combination of large distance + blending; R_sersic = 1.8" * 1.86 kpc / " = 3.35 kpc
    \end{tabular}
    \caption{{\bf QPE host galaxy properties} ordered by decreasing declination. Host coordinates are taken from PS1 or DES DR2. Bulge-total ratios (B/T) in the $g$-band, Sersic indices, and stellar mass densities are taken from \cite{2025ApJ...994..209G} when available, or from \cite{2011ApJS..196...11S} otherwise. BPT indicates the BPT classification of the nuclear narrow emission lines, when present (Q indicates quiescent / no emission lines visible). $u-r$ colors are taken from the SDSS catalogue, or computed from host SED modelling following \cite{2023ApJ...955L...6Y}. L$_{\rm ion, min}$ is calculated following \cite{2023ApJ...950..153F}. If multiple velocity dispersion ($\sigma_{\star}$) measurements exist in the literature, we provide the inverse variance weighted average. Black hole masses are estimated using the relation from \cite{2020ARA&A..58..257G} (full sample, including limits), where we provide the measurement uncertainty and the systematic (0.55 dex) uncertainty separately. Caveats of these estimates are discussed in the text.}
    \label{tab:host}
\end{sidewaystable}
\renewcommand{\arraystretch}{1.}

\subsection{Black hole and galaxy stellar masses}
Black hole masses of the QPE hosts are estimated using the M--$\sigma$ galaxy scaling relation of \cite{2020ARA&A..58..257G}. Velocity dispersion measurements ($\sigma$) are available for most host galaxies in the literature (except for AT2019vcb and AT2022upj) and are compiled in Table \ref{tab:host}; these are generally $<$100 km s$^{-1}$, indicating low mass black holes $\sim 10^{5-7}$ M$_{\odot}$ \cite{2022A&A...659L...2W, 2024ApJ...970L..23W}. For eRO-QPE5 and eRASSt J2344 we provide a new velocity dispersion measurement based on nuclear (1" aperture) MUSE spectra. We follow the methodology outlined in \cite{2017MNRAS.471.1694W}. For eRO-QPE5, we find $\sigma$ = 60$\pm$6 km s$^{-1}$, corresponding to log(M$_{\rm BH}) = 5.5 \pm 0.2$ (meas.) $\pm$ 0.55 (sys.). This is significantly lower than the independent mass estimate presented by \cite{2025ApJ...989...13A}. For eRASSt J2344, we find $\sigma$ = 106$\pm$10 km s$^{-1}$, corresponding to log(M$_{\rm BH}) = 6.6 \pm 0.2$ (meas.) $\pm$ 0.55 (sys.).

Note that for these low $\sigma$ values, typically $<100$ km s$^{-1}$, the M--$\sigma$ relation is not well calibrated and systematic uncertainties are significant. The relation from \cite{2020ARA&A..58..257G} incorporates a larger dynamic range in black hole mass and velocity dispersion than earlier calibrations, appropriate for low-mass systems such as QPE hosts. %However, we caution that the intrinsic scatter increases at the low-mass end of the relation, and the BH masses in this regime are often inferred from indirect methods with substantial observational limitations. 
Nevertheless, large uncertainties are associated with the M--$\sigma$ mass estimates and they should be treated cautiously; for example, self-consistent accretion disk modeling yields BH masses that can be significantly different (e.g. for eRO-QPE2 \cite{2025ApJ...980L...1W} and AT2019qiz \cite{2025arXiv251026774G}).

With these caveats in mind, it is clear that the QPE host black holes are smaller than those of the typical AGN population, spanning the range M$_{BH} \sim 10^{5-7}$ M$_{\odot}$.
These estimates are in line with the total galaxy stellar masses derived from SED fitting between 10$^{9.4-10.3}$ M$_{\odot}$ (see Table \ref{tab:host} and \cite{2025ApJ...994..209G}). This coincides with the black hole mass regime where tidal disruptions are observable and produce transient accretion disks, as black holes more massive than 10$^8$ M$_{\odot}$ would swallow a (solar-mass) star whole. %This overlap in BH and galaxy demographics reinforces the idea that QPEs share a connection to TDEs.}
%report typical galaxy masses within a narrow range between 10$^{9.4-10.3}$ M$_{\odot}$ (see Table \ref{tab:host}). Furthermore, \cite{2025ApJ...994..209G} showed through a comparison with a control sample matched in stellar mass and redshift that QPE host BHs are systematically under-massive compared to galaxies on the M$_{\star}$-M$_{BH}$ scaling relation, as has also been suggested for TDEs \cite{2022MNRAS.515.1146R}. 

\subsection{Star formation histories}
Several QPE hosts display properties characteristic of quiescent Balmer-strong (QBS) and post-starburst (PSB) galaxies \cite{2022A&A...659L...2W}, which are thought to have experienced significant bursts of star formation followed by rapid quenching within the past $\sim$0.1–1 Gyr. This is supported by spectroscopic diagnostics (strong H$\delta_A$ absorption in the absence of H$\alpha$ emission), photometric colors, and comparisons with stellar population synthesis (SPS) modeling.
SPS fitting with a combination of an old and a young population suggests that several systems have recently experienced a quenching episode, with the fraction of stars formed in the most recent burst comprising 1--15\% of the total stellar mass \cite{2024ApJ...970L..23W}.

The preference for QBS/PSB galaxies, combined with the blue nuclear colors, stellar concentrations and the high fraction of hosts found in the green valley, suggests that QPEs preferentially occur in galaxies that are rapidly transitioning from star-forming to quiescent galaxies, potentially following minor merger events. This points to an evolutionary stage in which the nuclear stellar density is high (e.g. \cite{2025arXiv251018985N}) and the rate of close encounters between stars and the central BH is enhanced, conditions that may play a key role in giving rise to QPEs.

\subsection{Emission line properties}
Emission-line diagnostics provide a complementary view of the nuclear environments of QPE host galaxies, offering insight into the presence and history of ionizing radiation sources. In particular, the relative strengths, spatial extents, and kinematics of emission line regions can distinguish between ongoing AGN activity, fading accretion episodes, and alternative ionization mechanisms.

\subsubsection{Nuclear emission lines}
Every QPE host galaxy has a nuclear narrow line region\footnote{The available SALT/RSS spectrum for eRO-QPE5 was reported not contain emission lines, but the MUSE data clearly shows H$\alpha$ and N\,\textsc{ii} emission.}. Evidence for a very compact but extended nuclear emission line region ($\lesssim$ 35 pc) has also been reported in HST narrow-band imaging of GSN 069 \cite{2024MNRAS.530.5120P}. Following a correction for stellar absorption, \cite{2022A&A...659L...2W} and \cite{2024ApJ...970L..23W} studied the line ratios to determine the dominant ionizing continuum and classify sources on the Baldwin-Philips-Terlevich (BPT, \cite{1981PASP...93....5B}) and W$_{H\alpha}$ versus [NII]/H$\alpha$ (WHAN, \cite{2011MNRAS.413.1687C}) diagrams (Table \ref{tab:host}). The majority of sources requires a hard ionizing continuum that cannot be generated by stars alone. The source of this radiation is typically assumed to be provided by an AGN, but emission from a hot TDE accretion disk is a viable alternative.

%\subsubsection{(Lack of) Broad line regions}
Deep optical (ground-based) spectroscopy does not show any evidence for persistent broad emission lines in the QPE host nuclei. This indicates that there is no standard broad line region, consistent with the nuclear WISE color cut typically used to identify AGNs (W1--W2 $<$ 0.8, Table \ref{tab:host} and \cite{2012ApJ...753...30S}) and the compact size of the accretion disks. The amount of extinction seen in the quiescent X-ray spectra is generally consistent with that expected from the Galactic contribution along the line of sight, indicating that very little neutral absorption occurs in the galaxy nuclei. 
%These findings indicate that there is no significant obscuring structure along our line of sight, disfavoring a type-2 AGN explanation for the emission line properties. 
%Furthermore, the quiescent X-ray spectra are very soft and lack the typical harder power-law emission that is ubiquitous (and often dominant) in AGNs.  

The absence of broad emission lines, together with the lack of significant nuclear obscuration and the ultrasoft X-ray spectra, strongly disfavors interpretations in which QPE hosts harbor a conventional AGN; instead, the observational evidence indicates that exploring alternative scenarios for the emission line properties is required. %Deeper limits on the presence of a BLR can be obtained with space-based narrow-slit spectroscopy, but the observational evidence indicates that exploring alternative scenarios for the emission line properties is required. 

\subsubsection{Extended emission line regions}
\label{sec:eelr}
In addition to long-slit spectra, spatially resolved (integral field) spectroscopy from the Mapping Nearby Galaxies at APO (MaNGA) survey with the BOSS spectrograph \cite{2015ApJ...798....7B}, and from the VLT/MUSE spectrograph, is available in published literature for 7 QPE host galaxies \cite{2023ApJ...950..153F, 2024ApJ...970L..23W, 2025ApJ...989...49X}. We incorporate results from 6 additional unpublished datasets here\footnote{MUSE observations for 2MASXJ0249 (PI: F. Bian, ID 111.24UJ), eRO-QPE4 (PI: T. Wevers, ID 113.26F6), ZTF19acnskyy \cite{2026arXiv260700921S}, AT2019vcb, AT2022upj, eRASSt J2344, and eRO-QPE5 (PI: T. Wevers, ID 117.2AW1).}, providing a complete sample of integral field spectroscopy for QPE host galaxies (Table \ref{tab:host}). 

\begin{figure*}
    \centering
    \includegraphics[width=0.45\linewidth]{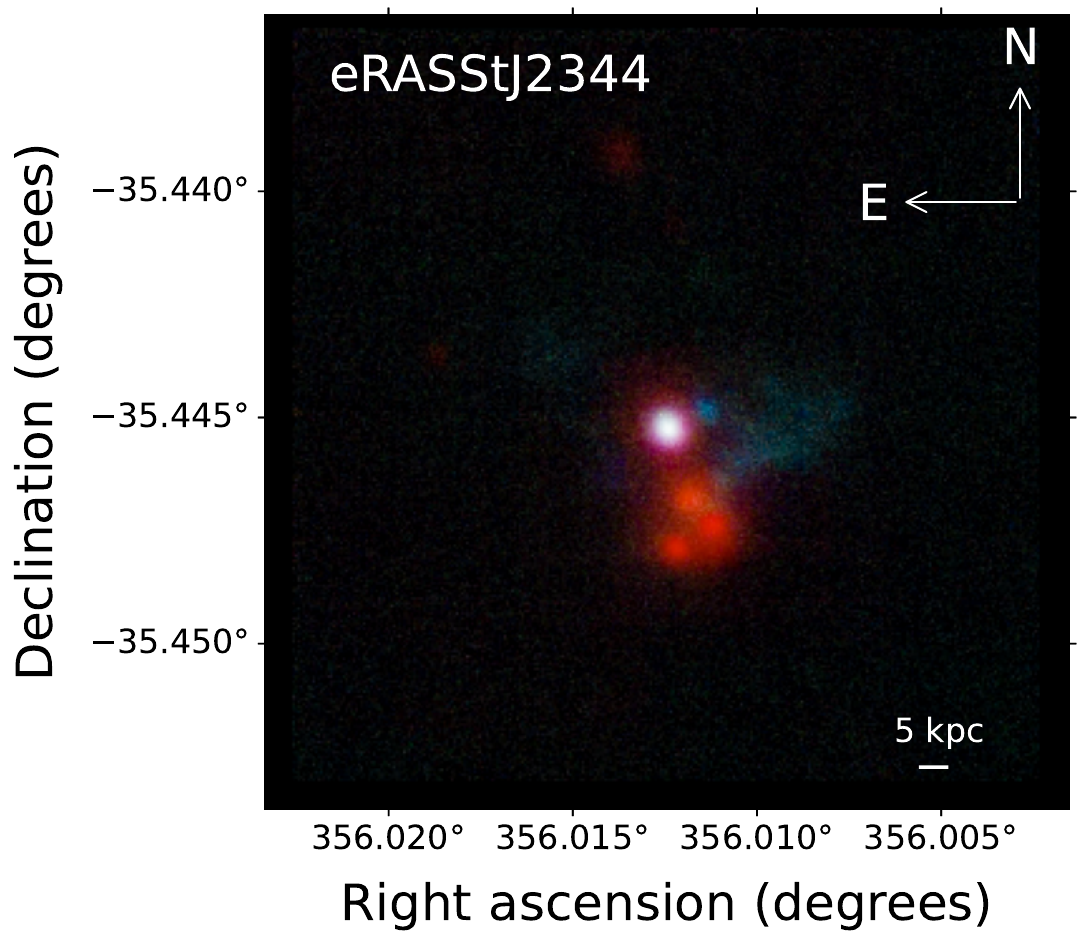}
    \includegraphics[width=0.45\linewidth]{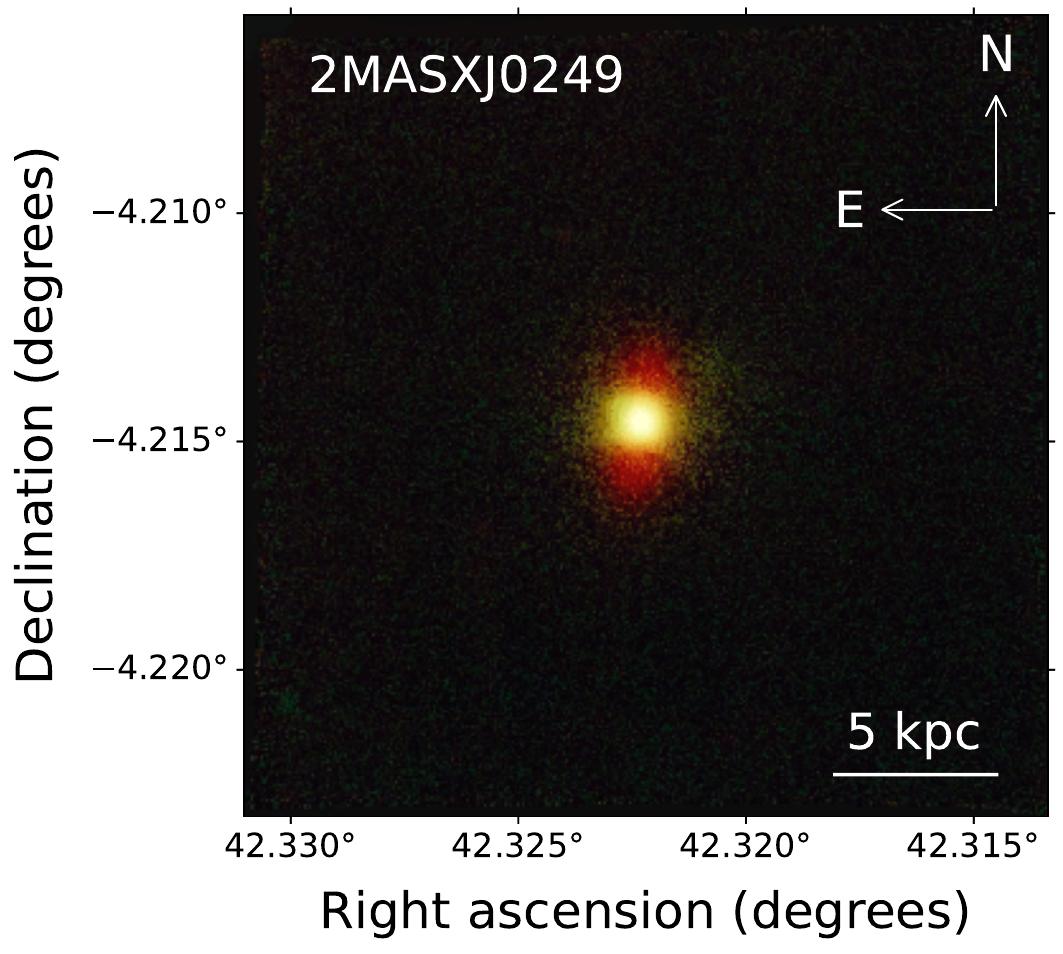}
    \includegraphics[width=0.45\linewidth]{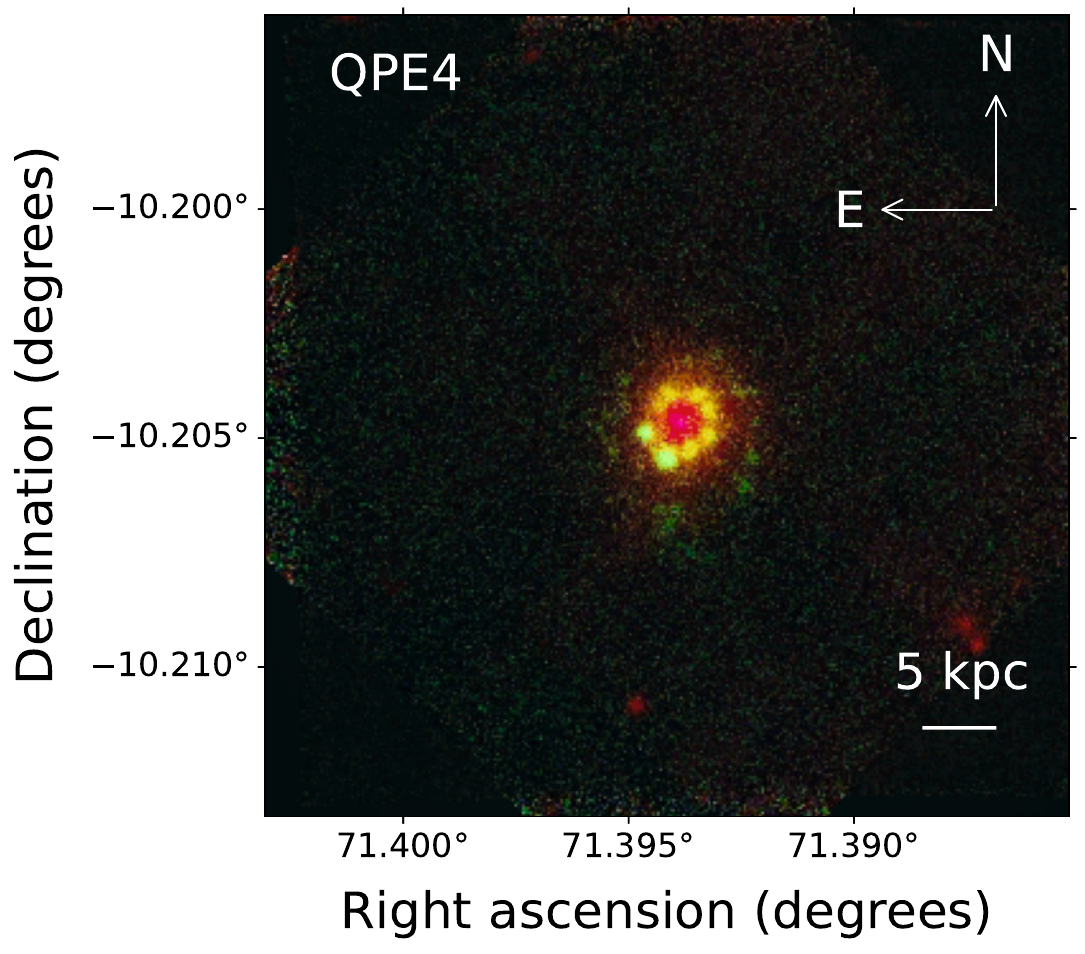}
    \includegraphics[width=0.45\linewidth]{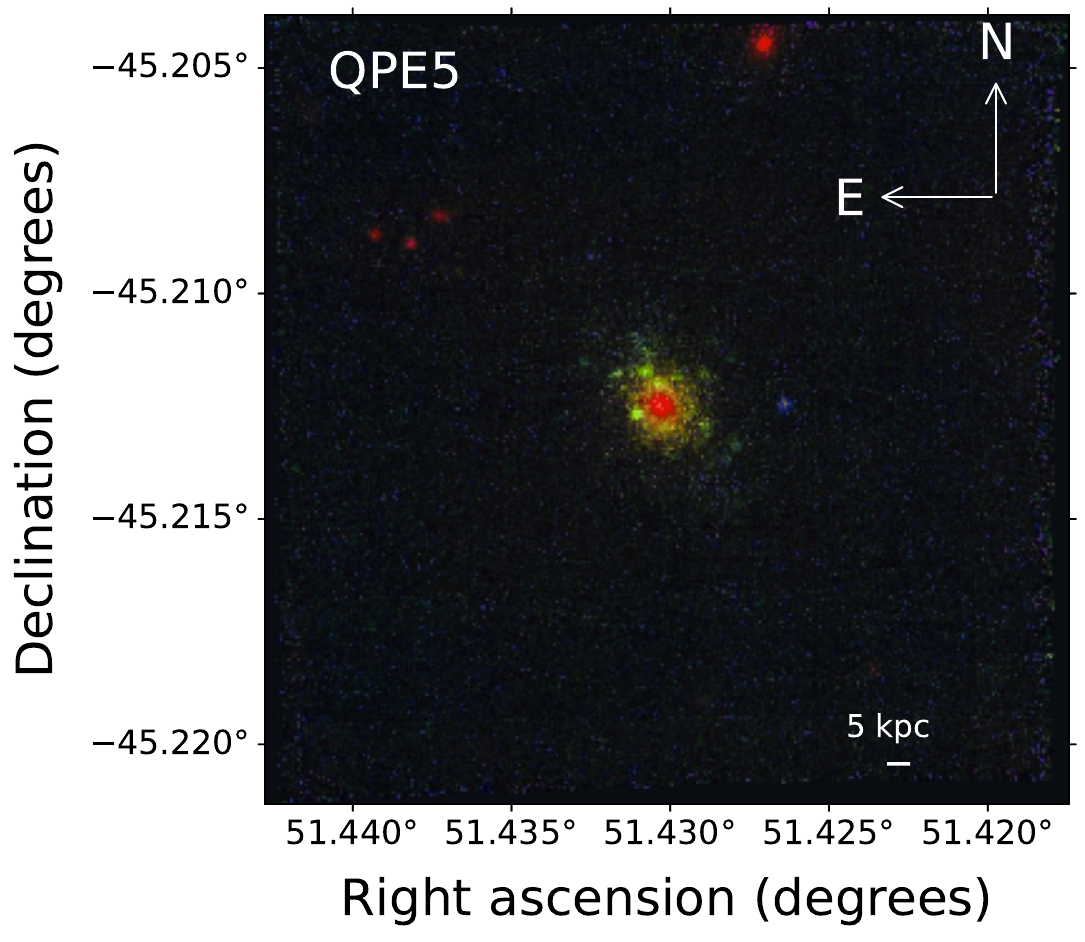}
    \caption{Three-colour images of the newly reported EELRs in the host galaxies of 2MASXJ0249, eRO-QPE4, eRO-QPE5, and eRASSt J2344. Red hues indicate the continuum emission around 8000 \AA; blue represents the O\,\textsc{iii} line emission and green represents H$\alpha$ emission. The 3 point sources located South of eRASStJ2344 are interlopers unrelated to the host galaxy. Image credit: J. Depasquale.}
    \label{fig:eelr}
\end{figure*}

The most striking discovery in these IFU data is the presence of extended emission line regions in 9/13 studied sources, namely GSN069, eRO-QPE2, RXJ1301 and AT2019qiz \cite{2023ApJ...950..153F, 2024ApJ...970L..23W, 2025ApJ...989...49X}, as well as 2MASXJ0249, eRO-QPE4, eRO-QPE5, eRASSt J2344 (reported in this work, see Fig. \ref{fig:eelr}), and Ansky \cite{2026arXiv260700921S}. This corresponds to a very high EELR fraction of $f_{\rm EELR}$ = 0.69$^{+0.31}_{-0.23}$ (following \cite{1986ApJ...303..336G} to calculate the 68\% confidence intervals), confirming previous findings \cite{2024ApJ...970L..23W} with a 2$\times$ larger sample size. We show the newly reported EELRs (2MASXJ0249, eRO-QPE4, eRO-QPE5, and eRASSt J2344) in Figure \ref{fig:eelr}; the EELR in eRASST J2344 is particularly large and spans $\sim$50 kpc in projected distance (from the NE component to the most westerly point).

%The other 5 EELRs are more irregularly shaped and have no clear spatial or kinematic connection to the nucleus (see \cite{2024ApJ...970L..23W}, and  This is more consistent with a different formation scenario, such as a minor merger, or the accretion or infall of gas from the circumgalactic medium. 
7 EELRs\footnote{The EELR in AT2019qiz has a bi-conical shape that appears to connect spatially to the galaxy nucleus \cite{2025ApJ...989...49X}, indicating that it is likely powered by (past) AGN activity. } have several striking properties which are atypical for EELRs found in other galaxies (see e.g. \cite{2020AJ....159..167L, 2023ApJ...950..153F} for large sample studies):
\begin{enumerate}
\item They have forbidden emission line ratios requiring the presence of a hard (non-stellar) ionizing continuum, and (projected) physical sizes up to 25 kpc from the nuclei. 
\item Their kinematics (both the velocities and widths) are decoupled from the stellar motions. Both the velocities and line widths are very low (generally $<$ 100 km s$^{-1}$), which strongly disfavors AGN-driven outflows as the origin of the gas.
\item The QPE energy budget is smaller (by orders of magnitude) than the the energy budget required to power these EELRs, and hence other emission mechanisms are required to explain their presence.
\item The present-day nuclear luminosity estimated from FIR observations (assuming the presence of an AGN) is insufficient to satisfy the EELR energy budget \cite{2024ApJ...970L..23W}. This implies that, if an AGN was previously present, it has faded significantly in luminosity over the last $\sim$10$^4$ years. Alternatively, long-lived TDE disks may provide the energy budget required to power the observed line luminosities and their ratios \cite{2024ApJ...970L..23W, 2025arXiv250314163M}.
\item The incidence fraction of EELRs in QPE hosts is much higher (overrepresentation of a factor $\gtrsim$10) than the typical EELR incidence in the general galaxy population \cite{2020AJ....159..167L}, AGN host galaxies \cite{2024MNRAS.530.1624K}, and even post-merger, post-starburst galaxies \cite{2023ApJ...950..153F}. It is similar to that of galaxies displaying tidal tails or companion interactions in continuum light \cite{2024MNRAS.530.1624K}, and tidal disruption event host galaxies \cite{2024ApJ...969L..17W}. 
\end{enumerate}

\cite{2025arXiv250314163M} recently used CLOUDY simulations to show that TDEs disks can match the required EELR energy budget, and hence that galaxies with elevated TDE rates are expected to have a higher incidence of EELRs provided that low density gas is present. Given that i) QPEs cannot satisfy the required EELR energy budget, and ii) the (largely) disparate selection functions for the QPE and known TDE samples\footnote{5/13 QPEs/QPE candidates (AT2019vcb, 2MASXJ0249, AT2019qiz, AT2022upj, and eRASStJ2344) were selected as TDEs (or TDE candidates) initially and then followed up or studied in more detail with X-ray observations / archival searches.}, this may suggest that QPE host galaxies also have elevated TDE rates.

These gas reservoirs were likely acquired through recent (minor) mergers or interactions, and their presence may influence stellar dynamics directly by exerting dynamical friction within the nuclear star cluster, while the interactions themselves may perturb stellar orbits. Both of these effects can enhance the rate of close encounters between stars and the central black hole.

%\footnote{Note that the reverse argument, that EELRs in PSB galaxies are the result of a high TDE rate, is unlikely, Starecheski et al. in prep. \cite{Starecheski}.}

\subsubsection{Transient coronal line emission}
Independent evidence for gas-rich nuclei comes from transient high-ionization emission lines. Two of the three QPEs discovered in the wake of optically-selected stellar disruptions (AT2019qiz and AT2022upj) display transient forbidden transitions such as [Fe,\textsc{vii}], [Fe,\textsc{x}], [Fe,\textsc{xiv}], and [S,\textsc{xii}], and are classified as extreme coronal line emitters (ECLEs; \cite{2024ApJ...977..258N, 2023MNRAS.525.1568S, 2025ApJ...983L..39C}). The appearance of these lines requires both a luminous, hard ionizing outburst and dense gas in the immediate vicinity of the nucleus. Such systems are rare among the AGN population but strongly linked to gas-rich galactic nuclei (\cite{2008ApJ...678L..13K, 2012ApJ...749..115W}), reinforcing the conclusion drawn from the EELRs that QPEs preferentially arise in nuclei harboring significant gas reservoirs, probed here on smaller (sub-pc to pc) scales than the kpc-scale EELRs.\\

Taken together, the characteristics of QPE hosts closely mirror those of TDE hosts \cite{2020SSRv..216...32F}, suggesting that QPEs arise in galactic nuclei where dynamical processes and transient accretion episodes are enhanced. 
%Any successful model for the origin of QPEs must therefore not only reproduce their distinctive X-ray phenomenology, but also explain their strong preference for this specific host galaxy demographic. 
%In \S \ref{sec:connection} we discuss the physical implications of these host galaxy properties in conjunction with the QPE properties in more detail.

% \bibliographystyle{style/spphys.bst}
% \bibliography{references.bib}

% \section{Physical Interpretation of the Observations}

\section{QPE origins: observational constraints and theoretical models}
\label{sec:models}
Despite the rapidly growing observational datasets, the physical origin of QPEs remains unknown. Nevertheless, the characteristic energetics, timescales, spectral evolution, and host galaxy properties of QPEs provide a set of increasingly stringent empirical constraints that any viable theoretical model must satisfy. Here we summarize these observational constraints in a model-agnostic manner before (briefly) assessing the extent to which proposed theoretical scenarios are consistent with the observed phenomenology. A detailed description of theoretical models is deferred to other Chapters.

% \subsection{Summary of observational constraints}
\subsection{Mass and energy budget}
The integrated energy output of QPEs places one of the most direct and model-independent constraints on their physical origin. 
QPEs show an integrated per-eruption energy output of $L\Delta t\sim 10^{45-48}$ erg. This can be converted into a crude constraint on the required mass budget, for some assumption about the radiative efficiency ($\eta$) of the mass to energy conversion. Adopting a range of $\eta\sim 10^{-3}-0.1$, one can infer a mass budget of approximately $M_{\rm QPE} \sim 10^{46-51}$ erg $\sim 10^{-8}-10^{-3}M_\odot$ per flare. Several known QPEs have exhibited continuous eruptions over a baseline of $\mathcal{O}(100-1000T_{\rm rec})$, approaching integrated mass budgets of $10^{-6}-1M_\odot$ over their observed lifetime. Moreover, the energy budget appears to increase for longer-period QPEs, primarily a consequence of the $t_{\rm dur}-t_{\rm rec}$ scaling increasing $L\Delta t$ for long-period sources. This is in contrast to the naive expectations for star-disk collision models, which generally predict a larger mass/energy reservoir at smaller disk radii and larger orbital velocities (e.g. \cite{2023ApJ...957...34L,2025arXiv250421456M}).

On one hand, the robust association with TDE disks in several QPE sources implies an available mass accretion reservoir of order $\sim0.5M_\odot$, suggesting that repeated accretion disk instabilities are a plausible scenario.  On the other hand, in EMRI models, where a companion interacts with a recently-formed accretion disk, the flare emission is powered by either the ejected disk material and/or mass lost from the orbiting stellar companion. Both models are consistent with the available mass/energy budgets in all but the most extreme cases. The inferred QPE lifetime for both disk instability and orbiter models is generally of order tens-hundreds of years. The characteristic blackbody radius inferred from the burst temperatures and luminosities corresponds to $\sim R_\odot$, a region which is much smaller than a typical accretion disk but is compatible with the inner disk in a disk instability model or the collision cross-section in an EMRI model.

\subsection{Timescales and spectral evolution}
In addition to energetics, the characteristic timescales and spectral evolution of QPEs provide powerful discriminants between competing physical models. The presently known QPE bursts release their energy over $\approx 1$ hr$-$few days. In comparison, the viscous timescale at 10$R_g$ around a $10^6 M_\odot$ SMBH is $t_{\rm visc} \sim 44$ hours, scaling as $\propto (\frac{R}{R_g})^{3/2}$ and linearly with $M_{\rm BH}$ for steady-state accretion disks (though see \cite{2025ApJ...992..114G} for modifications relevant to TDE disks which can introduce significant variation in $t_{\rm visc}$ even for fixed black hole mass). The shortest-duration QPEs (e.g. GSN 069, RX J1301, eRO-QPE2, eRO-QPE4) are unlikely to be powered by viscous accretion onto the SMBH given the short energy release timescale---although see \cite{2023ApJ...952...32P} for a mechanism to significantly reduce the accretion timescale---but in longer-period sources (e.g. eRO-QPE1, AT2019qiz, AT2022upj, ZTF19acnskyy) the possibility cannot be ruled out definitively. 

The same problem has been noted for the state transitions of changing-look AGN, akin to the thermal instabilities originally invoked for outbursts in accreting stellar-mass binaries. In fact, TDEs themselves are known to show X-ray variability significantly more rapid than the viscous timescale, with characteristic variability  spanning between the dynamical and thermal times of the accretion flows \cite{2026ApJ..1000...95C}. However, those rapid fluctuations occur only at the $\sim 10\%$ level, and it is uncertain whether the larger accreted mass budgets implied by QPEs can operate on the same sub-viscous timescales. Mechanisms for large-amplitude variability on the thermal timescale have been proposed for AGN \cite{2018MNRAS.480.3898N} and it is possible that related mechanisms could be relevant in QPEs. The observed rapid burst durations, particularly for short-period QPEs, strongly motivate scenarios in which the flare emission is powered by processes other than steady viscous accretion onto the black hole.

To this end, the $L-k_BT$ hysteresis pattern is one of the distinctive observational features which has been used to motivate models of the flare emission mechanism \cite{2025ApJ...983...40V}. Via analogy to the similar energy dependence of optical light curves observed in the rise/decline of Type Ia supernovae (e.g. \cite{2006ApJ...651..366K}), the harder-then-softer multi-band X-ray evolution of QPEs has been used to infer shock-powered emission in a homologously expanding envelope \cite{2025ApJ...983...40V}. Models which power the emission via shocks rather than accretion also provide a natural explanation for the short durations compared to $t_{\rm visc}$, conveniently fulfilling the timescale requirement even in short-period QPEs. 

The longer-duration eruptions are relatively more complicated to constrain, as one cannot straightforwardly rule out accretion-based models by the viscous timescale argument. Moreover, the energy budgets of QPEs appears to increase with growing recurrence time/duration---in contradiction to basic expectations for the mass budgets available at larger radii in TDE disks \cite{2025arXiv250421456M}---hence the longest-duration QPEs place the greatest strain on the relevant mass budget.

The burst recurrence timescales and their tight scaling with burst durations \cite{2025ApJ...989...13A} provide another model constraint. Among the most striking features of QPEs is their remarkable regularity, in stark contrast to the typically stochastic nature of X-ray variability from SMBHs, indicative of an underlying clock present in these systems. The sources show varying degrees of regularity: some QPEs (e.g. eRO-QPE2, GSN 069) show a long-/short-alternation pattern in recurrence time, with a difference of $\sim 10$\% and little scatter otherwise; other sources (e.g. eRO-QPE1, RX J1301, AT2022upj) show unpredictable scatter up to $\sim 50$\% between successive burst recurrence times. Any model to power the eruptions must also account for the roughly constant $\sim10-30$\% duty cycles due to the $t_{\rm dur}-t_{\rm rec}$ scaling, insofar that this is not driven by a selection bias (\S \ref{sec:biases}).

Taken together, the observed hysteresis patterns, burst durations, recurrence time quasi-regularity, and long-term timing variations suggest that QPEs are governed by a physical clock with some super-periodic structure, although the nature of this clock remains debated.

\subsection{Strengths and shortcomings of proposed scenarios}
We now briefly discuss the leading classes of theoretical models proposed to explain QPEs (further described in other Chapters), highlighting both the observational successes and the outstanding challenges faced by each scenario.
\subsubsection{Support for EMRI models}
EMRI models provide a convenient explanation for the relative stability of the underlying clock in QPEs, the long-/short alternation in the burst recurrence times \cite{2023A&A...675A.100F,2023ApJ...957...34L}, the observed hysteresis pattern in the bursts \cite{2023ApJ...957...34L,2025ApJ...983...40V}, and the event durations for the shorter-timescale bursts \cite{2025ApJ...993..186H}. They also offer a natural explanation for the super-periods tentatively observed on longer timescales in some QPEs by associating them with the precessional modes of the EMRI and/or accretion disk \cite{2023A&A...675A.100F,2024PhRvD.110h3019Z,2025ApJ...992..120C}, although it is not yet clear that such an association has been made uniquely. These models also provide a convenient explanation for the observed QPE rate $\sim10\times$ smaller than the TDE rate \cite{2024A&A...684L..14A,2025ApJ...983L..39C}, via arguments related to the lifetime of stellar EMRIs on $\gtrsim100R_g$ orbits around low-mass SMBHs \cite{2022ApJ...926..101M,2023ApJ...957...34L}. EMRI models which explain the emission via collisional shocks also provide a convenient explanation for the P-Cygni profiles seen in some QPE X-ray spectra, as they require the ejection of shock-heated debris undergoing homologous expansion over the course of the eruptions \cite{2025ApJ...983...40V,2025ApJ...993..186H}.

Collectively, these features make EMRI-based scenarios an attractive framework for explaining several defining properties of QPEs, including their regularity and short timescale bursts.

\subsubsection{Challenges for EMRI models}
Despite these appealing aspects, EMRI models also face significant challenges when confronted with the full diversity of observed QPE behavior. These include the unexpectedly large energetics of long-duration eruptions \cite{2025arXiv250421456M}, which are difficult to reconcile with the mass budgets expected from star-disk collisions or stellar stripping at larger orbital radii. Moreover, early attempts at uniquely associating the super-periodic modulations with EMRI precessional modes have faced some difficulty: the clear prediction from EMRI models undergoing relativistic apsidal precession is that even and odd bursts should undergo sinusoidal modulations in anti-phase, in contrast with observations in GSN 069, eRO-QPE1, and eRO-QPE2 \cite{2025A&A...693A.179M,2024ApJ...965...12C,Arcodia2026}. 

Additionally, eRO-QPE1 and ZTF19acnskyy display oscillation timescales of $\sim6\times$ and $\sim16\times$ the QPE recurrence period, which is more rapid than can be easily explained by general relativistic precessional modes. A possible explanation that has been invoked is rapid disk precession \cite{2023A&A...675A.100F,2024PhRvD.110h3019Z,2024ApJ...965...12C}; however, it is uncertain whether newly-formed TDE disks such as those relevant for QPEs can sustain rigid-body precession over the several-year lifetime during which QPEs have been observed \cite{2023A&A...675A.100F}. Some studies have suggested the effect of a third body in a hierarchical SMBH binary to explain both the correlated even/odd residuals and variation timescales and amplitudes \cite{2025A&A...693A.179M,Arcodia2026}; however, this model requires a fine-tuned combination of components, warranting further investigation.

It is not yet clear how to resolve these tensions, and it may be that reproducing QPE timing phenomenology will require significant modifications to the EMRI picture. Moreover, hydrodynamical simulations of star-disk collisions face some difficulty in producing the observed event durations for longer-timescale QPEs. They also produce unexpectedly asymmetric ejecta on either side of the disk, in contrast to the comparable luminosities of even/odd bursts in GSN 069 and eRO-QPE2 \cite{2025ApJ...993..186H}. Nevertheless, EMRI models grant some flexibility in the prescription of the emission mechanism, response of the stellar-mass companion, and the orbital configurations; these particular areas warrant detailed follow-up studies in light of the above tensions.

\subsubsection{Disk instability models}
Given the relatively larger uncertainty and degrees of freedom in our understanding of SMBH accretion processes, disk instability models are both more flexible in their ability to accommodate observations, and more difficult to falsify or verify via specific observational predictions. There have been several formulations of disk instability models involving magnetic pressure instabilities in super-Eddington TDE disks \cite{2023MNRAS.524.1269K}, thermal-viscous instabilities akin to the disk instability model invoked in cataclysmic variables/X-ray binaries \cite{2022ApJ...928L..18P}, disk tearing (e.g. \cite{2021ApJ...909...82R}), or precession of super-Eddington flows \cite{2025MNRAS.537.1688M}. One potential analogy is provided by the ``heartbeat'' states seen in some X-ray binaries \cite{2000A&A...355..271B, 2011ApJ...742L..17A}: large-amplitude, quasi-regular flaring cycles attributed to radiation-pressure-driven limit cycles in the inner disk, with burst profiles and duty cycles qualitatively similar to QPEs. They demonstrate that
disks can sustain structured, repetitive outbursts rather than purely stochastic variability, although these heartbeat states appear more irregular in comparison to the QPE phenomenon.

By construction, disk instability models are capable of reproducing the peak temperatures, spectral hysteresis, event durations, mean recurrence times, and in some cases, the irregularity in recurrence times \cite{2025MNRAS.537.1688M,2025ApJ...989..196P}. They have not yet been extended to the point of examining super-periodic timing variations, and do not immediately provide an explanation for the observed QPE rate. In light of the TDE association of QPEs, it is not clear why \textit{some} TDE disks would exhibit such dramatic regular changes while they generally show stochastic X-ray variability \cite{2026ApJ..1000...95C}. Disk instability for QPEs thus may require an fine-tuned configuration, though recent work in \cite{2025ApJ...989..196P} found the peak luminosities and recurrence times in disk instabilites are sensitively dependent on the mass accretion rate, providing a partial explanation for both the apparent fine-tuning as well as the dramatic period/luminosity variations seen in e.g. GSN 069 and Ansky \cite{2023A&A...674L...1M,Chakraborty2026}. In general, disk-instability scenarios remain viable but lack a clear set of unique, testable predictions that would distinguish them decisively from orbiter-based models.\\

In summary, existing observations impose strong constraints on the physical origin of QPEs, yet no single model currently provides a complete explanation for the full range of observed properties. Progress will hinge on identifying observations capable of breaking remaining degeneracies; particularly powerful will be those that probe geometry, orbital dynamics, or the coupling between transient accretion flows and their environments.

% \section{Connection to Other Phenomena}

\setcounter{section}{6}
\section{Connections and overlap with other phenomena}
\label{sec:connection}

%While QPEs constitute a distinct class of nuclear variability, placing them in the broader landscape of transient and quasi-periodic phenomena can provide a broader physical context, helping to identify shared timescales, environments, and emission mechanisms (as well as key differences). We assess such connections critically with the goal of clarifying which aspects of QPE behavior are likely generic to accretion physics and which appear to be unique.

As described earlier in this Chapter, a strong connection to TDEs is motivated by temporal coincidences and similar host galaxy demographics. More recently, parallels have also been drawn to quasi-periodic oscillations (QPOs) observed in accreting black hole systems \cite{2025Natur.638..370M}.%, and to quasi-periodic outflows (QPOuts) reported in some accretion-powered transients \cite{2024SciA...10J8898P, 2024arXiv241012090Z}. 
We assess the evidence linking QPEs to other forms of episodic black hole activity, and discuss what these comparisons reveal about the processes driving their recurring, high-amplitude X-ray flares.

\subsection{Tidal disruption events}
Although QPEs are a distinct phenomenon from TDEs, a growing body of evidence suggests that they are closely linked. Because the number of known QPEs is small, and discovering new events remains challenging, one way forward is to leverage host galaxy properties, long-term variability, and circum-nuclear environments (in addition to spectral/timing characteristics) to obtain a better understanding of their physical origins.

QPEs have been observed in several galaxies in the months--years following a TDE flare classified through optical or X-ray spectroscopy (e.g. \cite{2024Natur.634..804N, 2025ApJ...983L..39C, 2025A&A...697A.159G}. Several other QPEs display long-term declining and/or rebrightening X-ray lightcurves, consistent with the tail of a past, unseen tidal disruption \cite{2019Natur.573..381M, 2024A&A...684A..64A}. Given the rarity of both classes of events, this temporal association provides direct evidence that QPEs and TDEs are interconnected. ZTF19acnskyy is associated with a major accretion flare that is not definitively a TDE---though see \cite{2025ApJ...994L..16Z}---implying that the link may more broadly reflect the formation of a new accretion flow rather than TDEs specifically.

Taken together, the temporal associations, compact accretion disks and shared host demographics ($\S$ \ref{sec:host}) all point towards a close physical connection between QPEs and TDEs, even if the physical implications are not yet fully understood.

Note that, as described further in Sec. \ref{sec:biases}, it will be important to keep in mind detection biases (such as the dedicated follow-up of TDEs in search for more QPEs) when analyzing QPEs in the broader context of transient phenomena.

\subsection{Quasi-periodic oscillations}
Quasi-periodic oscillations (QPOs) provide a useful point of comparison because they operate on similar timescales in some accreting black hole systems. 
%However, QPOs and QPEs differ fundamentally in their amplitude, duty cycle, and light-curve morphology, indicating that they likely arise from distinct physical processes or geometries.

QPOs are a phenomenon associated with the dynamical timescale of accreting black holes, which is manifested by a narrow excess in power in the Fourier spectra \cite{2016AN....337..398M,2019NewAR..8501524I,2000ARA&A..38..717V}. QPOs have been studied in detail for stellar mass black holes, operating on timescales from milliseconds to seconds.    
QPOs associated with SMBHs operate at lower frequencies, from minutes to days and potentially years for the most massive SMBHs \cite{2016AN....337..417A}, which partially overlaps with QPE timescales (hours-days).

The fundamental (observational) differences between QPOs and QPEs is in their amplitudes, duty cycles, and light curve morphology. Specifically, for QPEs the amplitude is larger, and the emitting phase is short in comparison with the recurrence timescale and hence the duty cycle is smaller than unity ($\sim 0.1-0.3$; burst-like behaviour). For QPOs, the behavior is more sinusoidal with lower amplitude, and the emitting phase covers the whole recurrence timescale such that the duty cycle approaches unity ($\sim 1$).

One peculiar example is 2XMM J123103.2+110648 with a low-mass SMBH ($\sim 10^5\,M_{\odot}$) that exhibits significant QPO-like variability with a period of $\sim 3.8$ hours. 
The X-ray spectrum of the latter source is ultrasoft, consistent with pure thermal or Comptonized disk emission, and the rms amplitude of the oscillations increases with energy \cite{2013ApJ...776L..10L}. 

The pattern of periodic flares seen in the XMM-Newton light curves was reanalyzed by \cite{2023MNRAS.518.3428W} to assess whether they are more consistent with smooth sinusoidal QPOs or burst-like QPEs. These authors found that the pattern could be transitional, i.e. changing from epoch to epoch, hence J1231 could be a relevant source for follow-up studies of the potential QPO-QPE connection.
%The key open questions are whether there exists a physical connection between these two phenomena, and whether transitions between QPEs and QPOs can occur under certain conditions and on what timescales.

%One intriguing example is the recently observed mHz QPO in 1ES~1927+654 \cite{2025Natur.638..370M}. Over a two-year period, a narrow, quasi-periodic signal was detected, with a recurrence time that decreased from 18 minutes to 7.1 minutes. %The smooth evolution of the timescale and the soft, thermal X-ray spectrum suggest a link to the dynamical timescale of the inner accretion flow. However, the inferred emission radius lies very close to or even within the ISCO for the estimated black hole mass, challenging standard interpretations in terms of orbiting bodies or resonances. 
%Unlike typical QPEs, the variability amplitude remained modest and the emission was quasi-sinusoidal, consistent with high duty-cycle QPO behavior. Nonetheless, the source may represent a system with the same underlying driver as QPEs (e.g. disk instabilities or inspiraling bodies, \cite{2025Natur.638..370M}) but manifesting differently due to specific system parameters such as the nature of the orbiting body, orbital geometry, or viewing angle.

\subsubsection{QPO-like signals in QPE sources and their interpretation}

QPO-like variability has been reported in the quiescent phases of two QPE sources. In GSN~069, low-intensity modulations appear in the residuals of the folded soft X-ray light curves \cite{2023A&A...674L...1M, 2023A&A...670A..93M},
with a recurrence time comparable to the QPEs and a peak delayed by several 1000~s with respect to the eruptions. Notably, this signal is present only during epochs when the eruptions are bright and regular, suggesting that it is linked to the QPE-driving mechanism itself rather than constituting an independent phenomenon. In RX~J1301.9+2747, a tentative $\sim$1500~s periodicity was reported in quiescence \cite{2020A&A...644L...9S}, an order of magnitude shorter than the $\sim13\,000-20\,000$~s QPE recurrence time \cite{2020A&A...636L...2G}.

The coexistence of burst-like QPEs and low-amplitude QPO-like variability within the same sources raises the question of whether both reflect different modes of a common physical framework, with the duty cycle of energy release as the distinguishing parameter. 

Non-linear limit cycles of accretion disks can generate both eruptive and quasi-sinusoidal variability depending on disk conditions and magnetic stresses \cite{2000ApJ...542L..33J, 2002ApJ...576..908J, 2006MNRAS.372..728M, 2023MNRAS.524.1269K}, in which case the QPE/QPO phenomenology reflects different regimes of the same instability rather than an external clock (see also \cite{2025Natur.638..370M}).

Alternatively, in orbiter models transitions between the two regimes can arise through secular evolution of the orbital parameters \cite{2021ApJ...917...43S, 2023MNRAS.523L..26K, 2023MNRAS.526...69T} or through changes in how the dissipated energy is radiated, from impulsive disk-crossing shocks to more gradual dissipation via ablated or stripped material \cite{2025ApJ...978...91Y, 2025ApJ...991..147L}. 

For example, in the model of \cite{2023MNRAS.523L..26K}, a mass-losing white dwarf on an orbit that circularizes and shrinks through gravitational-wave losses eventually deposits matter that evolves on the viscous timescale, smearing the burst-like variability toward a duty cycle of unity. Some AGN QPOs could then represent a later evolutionary stage of QPE-hosting nuclei \cite{2023MNRAS.518.3428W, 2023MNRAS.523L..26K, 2023A&A...670A..93M}. \\

At present, no definitive physical connection between QPEs and QPOs has been established. Progress can be made through systematic searches for sources with intermediate or evolving variability, and long-term monitoring to track changes in duty cycle, coherence, and spectral properties.

\setcounter{section}{7}
\section{Challenges, opportunities, and the road ahead}
Having reviewed the phenomenology of QPEs, their host environments, and consistency with theoretical models, we now turn to some of the key challenges and opportunities that will define progress in the coming years.

\subsection{Challenges}
QPEs are a recently discovered phenomenon that requires expensive observational campaigns to uncover. Their physical origin(s) remains unknown to date, but they have the potential to provide unique probes of stellar dynamics in galactic nuclei, accretion disk physics, and potentially even multi-messenger astronomy. We summarize some of the main obstacles the field faces, and avenues the community could pursue to tackle them. Given the diverse and heterogeneous properties of the known sample, we identify increasing the sample size as a top priority. This will allow characterizing the biases inherent to the existing systems and robustly map out QPE properties at the population level. Finding more QPE systems faces many challenges:

\begin{itemize}
    \item \textbf{Limited instrumental energy coverage:} %All known QPEs have been detected via soft X-ray emission (0.1--1.5 keV), with typical blackbody temperatures of 50--200 eV.  
    With a temperature of 100 eV and neglecting the effects of absorption, $\sim20\%$ of the energy flux of the QPE emission is below 0.2 keV when the source is at $z=0$. This fraction rises to 60\% at $z=1$. This limits detections to nearby sources and instruments with strong low-energy response. Future sample size growth will likely be limited due to the lack of a sensitive, wide-field X-ray survey mission. 

    \item \textbf{Short-lived and low-duty cycle flares:} QPEs are short-lived, low duty cycle phenomena. The detection probability scales with the sum of source duty cycle and observing duty cycle. With limited time available for monitoring, observers must find a balance between short, dense monitoring or sparse, longer-baseline coverage to maximize the discovery probability of new QPEs. 

    \item \textbf{Interpretation of non-detections:} The absence of eruptions may be observational (missed due to cadence/gap) rather than intrinsic (QPE turned off). For candidates identified in archival data with few cycles, a single non-detection may dissuade follow-up.

    \item \textbf{Poorly constrained long-term evolution:} Only a handful of QPEs have been monitored over multiple years. In some cases (e.g., GSN 069, eRO-QPE1), flare timing and strength evolve significantly, or eruptive activity ceases entirely. Understanding QPE lifetimes and long-term variability may ultimately provide some of the strongest constraints on theoretical models.

    \item \textbf{Fragmented follow-up strategies:} Most follow-up campaigns focus on individual sources that are largely uncoordinated. Long-term multi-facility coordination including both high cadence monitoring and deep stares present a significant logistical challenge.

    \item \textbf{Multi-wavelength counterparts:} Most models (understandably) are focused on X-ray emission, but coordinated UV, optical, and radio campaigns have the potential to reveal faint or transient counterparts (e.g. \cite{2024ApJ...963L...1L}); QPE-correlated UV variability has been reported in 1 source \cite{2026ApJ..1000L..57G}, potentially offering new constraints on emission geometry, environment, and outflows.\\

\end{itemize}

% \begin{figure}
%     \centering
%     \includegraphics[width=0.7\linewidth]{Figures/Section8/QPE_DutyCycle.pdf}
%     \caption{Illustration of an observational constraint of QPEs: in order to ensure the detection of an eruption with certainty, the sum of the duty cycles of the eruptions and of the observations in one period must be over 1.}
%     \label{fig:dutycycle}
% \end{figure}

\subsection{The X-ray landscape and prospects for the field}
The landscape of X-ray observatories will play a central role in shaping the progress of QPE discovery and characterization. Their value lies in two complementary observational modes: discovery and follow-up. For the former, an all-sky X-ray survey would be most efficient to obtain an unbiased sample (as was demonstrated by eROSITA). For the latter, a combination of (1) deep, continuous exposures that capture multiple eruptions with high sensitivity, and (2) high-cadence monitoring over extended baselines to constrain long-term behavior would be ideal. 

While not optimized for soft X-ray sensitivity, missions such as Einstein Probe, SVOM, and eXTP can provide crucial long-term monitoring and follow-up capabilities in the next 5 years. This will bridge the gap to several proposed X-ray observatories for the early-mid 2030s that would directly enhance QPE science, including eXTP \cite{2025SCPMA..6819502Z} and NewAthena \cite{2013arXiv1306.2307N, 2025NatAs...9...36C}. 

In light of the limited prospects for an X-ray sky surveyor, we expect archival searches to remain a productive avenue for discovering new QPEs. Promising directions include more systematic searches in soft X-ray archives (e.g. \cite{2025arXiv251122520Q}), and searching for QPEs with harder spectra or atypical duty cycles. 
In addition, targeted TDE monitoring has proven to be an effective way to find new QPE sources; this can be particularly complementary to find sources that fall outside of the traditional QPE definition.

\subsection{Multi-wavelength prospects}
Multi-wavelength observations offer several avenues for constraining their physical environments and broader nuclear context. While most theoretical models focus on thermal X-ray emission, some predict the presence of harder components or signatures that may emerge in other bands (see \cite{2026ApJ..1000L..57G} for an example). 

Several upcoming and planned facilities could help test these predictions, either directly or by providing opportunities to increase the QPE sample size. 

ULTRASAT will offer high cadence wide-field UV monitoring ($\sim 200\,{\rm deg^2}$, NUV (250 nm) photometry down to $\sim 21$ mag, \cite{2024ApJ...964...74S}), potentially detecting quiescent disk emission in QPE hosts, and providing strong constraints on the presence of UV QPEs \cite{2024ApJ...963L...1L}, although the poor expected spatial resolution may limit the sensitivity of such searches in practice, especially if hot TDE-like disks are ubiquitous among QPE sources. Other planned UV missions will be beneficial in this effort by increasing the monitoring cadence and measurements in other photometric bands (e.g. the two-band NUV-FUV photometry mission QUVIK, \cite{2024SSRv..220...11W,2024SSRv..220...29Z,2025JATIS..11d2222Z}), or by obtaining deeper and higher spatial resolution UV imaging at lower temporal cadence (UVEX, \cite{2021arXiv211115608K}). 
%{\bf The case of the Ansky source with the UV variability correlated with QPEs indicates that longer-period QPEs may be better suited for searching for UV counterparts. This is due to the temporal smearing of the UV response that is more profound at shorter periods. In general, it is not yet clear whether the QPE-correlated UV variability operates on the diffusion timescale within the expanding shocked blob or whether it takes place on the light-crossing timescale within the X-ray irradiated accretion disk \citep{2026ApJ..1000L..57G}.} 
Similarly, high cadence optical surveys such as the Argus Array \cite{2022PASP..134c5003L} can provide time-resolved limits to the optical emission associated with QPEs. 

The Vera Rubin observatory, Argus Array and other optical wide-field surveys will vastly increase the sample of optical TDEs, some of which may evolve into QPE-producing systems. However, the redshift range does not always overlap with the capabilities of current X-ray instruments to discover QPEs, which may limit these synergies (e.g. the median expected redshift of Rubin-discovered TDEs is $z \sim 0.7$, likely too high to detect large numbers of QPEs). 

In the IR, the Nancy Grace Roman Space Telescope may detect late-time infrared echoes from past flares, which have been suggested as a possible avenue to further optimize searches for candidate QPE systems following TDEs \cite{2025arXiv250212078P}, and offering a window into circumnuclear dust geometry and flare energetics. 

At radio frequencies, hints of variability have been reported but follow-up observations either lack the sensitivity or the temporal resolution to provide unambiguous evidence. So far the only significant radio emission associated with the soft X-ray flare was measured for the repeating nuclear transient Swift J0230+28 \cite{2024NatAs...8..347G}, whose properties, such as the temperature evolution during eruptions, deviate from those of QPEs. Future radio facilities including the Deep Synoptic Array \cite{2019BAAS...51g.255H} will provide the increased sensitivity ($<$1 $\mu$Jy) necessary to detect or constrain compact outflows or past jet activity.

Coordinated observations across the electromagnetic spectrum provide a promising path toward constraining the emission geometry, disk structure, and possible jet or outflow components in QPE systems.

\subsection{Multi-messenger prospects}
\label{sec:multimessenger}
If QPEs are associated with extreme mass ratio inspirals (EMRIs), they could serve as unique joint GW--EM laboratories. The two channels are complementary: EM-only estimates of the SMBH mass and spin are sensitive to the adopted mass-determination method and
accretion disk model (surface density, scale height, optical depth, viscosity), which vary across the QPE sample with accretion rate, leaving degeneracies in light curve and spectral fits (e.g. between companion mass and disk surface density, \cite{2025arXiv250807961L}). A GW detection would constrain the orbital dynamics uniquely (SMBH mass and spin, companion mass, eccentricity, inclination, and luminosity distance) while the EM data provide the redshift and orbital period. 

The realization that QPEs may be EM counterparts to GW sources (e.g. \cite{2021ApJ...917...43S, 2024MNRAS.532.2143K, 2025PASJ..tmp..141S, 2025arXiv250807961L}) has stimulated calculations of the expected GW signal, with detection prospects set mainly by the mass of the orbiting body and its distance from the SMBH. The EMRI merger timescale \cite{1964PhRv..136.1224P} (assuming two QPEs per orbit) determines whether a known system might merge within the LISA timeframe:
\begin{align}
    \tau_{\rm merge} &\simeq \frac{5c^5}{256 \pi^{8/3}G^{5/3}}\frac{P_{\rm QPE}^{8/3}}{m_{\rm EMRI}M_{\bullet}^{2/3}}\,,\notag\\
    &\simeq 2.9 \times 10^6\,\left(\frac{P_{\rm QPE}}{10\,{\rm hours}} \right)^{8/3} \left(\frac{m_{\rm EMRI}}{1\,M_{\odot}} \right)^{-1}\left(\frac{M_{\bullet}}{10^6\,M_{\odot}} \right)^{-2/3}{\rm yr}\,
    \label{eq_merger_timescale}
\end{align}
This is too long for LISA in the 2030s--2040s for fiducial parameters, although an
intermediate-mass black hole orbiter ($m_{\rm EMRI}=1000\,M_{\odot}$,
$P_{\rm QPE}=1$ hour) would yield $\tau_{\rm merge}\sim 6.3$ yr. Conversely,
backward propagation of the orbital frequencies of sources merging during LISA
operations (2037--2041) places the present-day QPE frequency band at
$\nu_{\rm QPE}=0.46 \pm 0.22$ mHz, i.e. periods of $\sim 0.41-1.16$ hours
\cite{2024MNRAS.532.2143K}, motivating searches at the short-period tail of
the recurrence time distribution.

Directly detecting any of the currently known QPE sources with LISA will be challenging \cite{2024MNRAS.532.2143K, 2024A&A...690A..80A, 2025arXiv250807961L} (see also Fig.~13 in \cite{2025ARA&A..63..379K}): their implied EMRIs peak at frequencies $<$1 mHz, where LISA is substantially less sensitive than the 1--100 mHz range. Multi-messenger QPE studies will likely require future observatories such as TianQin \cite{2016CQGra..33c5010L} and Taiji \cite{2024SCPMA..6720412J}, or detectors sensitive in the $\sim\mu$Hz regime \cite{2021ExA....51.1333S}.  \\

In addition to gravitational wave emission, studies targeting other messengers including neutrinos are also starting to emerge, although similar to GWs, this is strongly model dependent. \cite{2025PhRvD.112f3031M} study the neutrino emission in star-accretion disk interactions. In this scenario hadrons (protons) can be accelerated to energies $>10\,{\rm TeV}$, and the neutrinos (with energies $<$ 10 TeV) are produced via $pp$ and $p\gamma$ interactions in the star-disk shock. Since these calculations are strongly model dependent, they should be regarded as tentative until the underlying nature of QPEs is more firmly established. 

%Furthermore, there is a likely association of a high-energy neutrino ($\sim 0.2–0.3$ PeV) detected by IceCUBE with the TDE AT2019dsg \cite{2021NatAs...5..510S}, coinciding with the phase following the optical peak when the radio and X-ray emission implied the presence of a mildly relativistic outflow. This provided a favourable site for proton acceleration in the dense radiation field and hence for the $p\gamma$ interaction, producing neutrinos. Since there is a growing evidence for the association of some QPEs with past TDEs, the example of AT219dsg shows that neutrinos can also be produced during the post-TDE evolution independent of QPEs, and in particular the tentative star-disc crossings that could be responsible for them. 

\section{Summary}
The discovery of QPEs has revealed a new window of opportunity to study black hole accretion processes and their environments, including the fundamental physics governing the evolution of accretion disks and stellar dynamics in galactic nuclei, and potentially the EM counterparts to low frequency gravitational wave sources. 

The known sample has grown to 13 systems to date, revealing a surprising degree of coherence in aspects including their spectral properties and duty cycles, and the connections to post-merger and green-valley host galaxies. At the same time, their timing properties and long-term evolution show a wide variety of behaviors, challenging the existing theoretical frameworks aiming to explain their origin as a phenomenon.

Over the next five years, addressing several key observational limitations will be essential to advancing our understanding of these sources. The short duty cycles and modest luminosities make them difficult to detect serendipitously, and current monitoring campaigns are constrained by cadence, sensitivity, and (X-ray) mission availability. As a result, the known sample likely represents a biased subset of a larger underlying population.

We conclude by outlining a number of open questions whose answers are likely to inform both the physical origin of QPEs and the design of future theory-driven observational efforts.

\begin{enumerate}
    \item How do we define the observational boundaries of the QPE phenomenon? Are there new phenomena or multi-wavelength counterparts which have not yet been observed?
    \item In the absence of a sensitive wide-field X-ray survey mission, how do we increase the discovery rate of QPEs without biasing the sample (e.g. towards only TDE-associated sources)?
    \item Is there a single theoretical model that can explain the QPE population, or does the heterogeneity in observed behavior point towards multiple sub-classes?
    \item How and why do burst properties evolve over months to years? What causes QPEs to fade, stop, or resume, often with different properties? What is the typical lifetime of a QPE source?
    \item What is the true rate of QPEs in the local universe, and how do current selection effects bias our view?
    \item Why do only some post-TDE systems exhibit QPEs, and what distinguishes them from non-QPE-producing TDEs?
    \item What is the physical origin of the super-periodic variations observed in the burst timings of several QPE sources?
\end{enumerate}

As a newly identified phenomenon, the properties and behavior of QPEs remain to be charted in full. Their discovery has opened a window onto processes that operate on accessible timescales and probe fundamental questions about black hole fueling, accretion disk evolution, and variability. 
%As observations and models evolve together, QPEs provide a new framework for studying the inner workings of galactic nuclei.

% \section{Summary and Outlook}
%\subfile{9_summary}

% \section{Summary and Outlook}
% Recap of key observational findings.\newline
% Open questions and how observations might answer them.\newline

\begin{acknowledgement}
We are grateful to (in alphabetical order) R. Arcodia, F. Bian, M. Masterson, B. Mockler, M. Newsome, J.J. Ruan, P. Sánchez-Sáez, Y. Yao, and Z. Zhang for providing comments, suggestions, data or analysis tools. We warmly thank J. Depasquale for creating the RGB composite images of the EELRs shown in Figure \ref{fig:eelr}.
We thank the two anonymous referees for thoughtful comments that improved the manuscript.

Based on observations collected at the European Southern Observatory under ESO programs 111.24UJ, 113.26F6, and 117.2A1W. This work includes data gathered with the 6.5 meter Magellan Telescopes located at Las Campanas Observatory, Chile.

MZ acknowledges the support of the Czech Science Foundation Junior Star grant no. GM24-10599M.
MG is funded by Spanish MICIU/AEI/10.13039/501100011033 and ERDF/EU grant PID2023-147338NB-C21.

\end{acknowledgement}

\bibliographystyle{spphys.bst}
\bibliography{references.bib}
\end{document}